\documentclass[12pt]{article}

\usepackage[verbose]{scicite}
\usepackage{graphicx}
\usepackage{times}
\usepackage[margin=1.3in]{geometry}
\usepackage{caption}
\usepackage{threeparttable}
\usepackage{booktabs}
\usepackage{hyperref}
\usepackage{endnotes,etoolbox}
\usepackage{subcaption}
 \usepackage{mwe}
 \usepackage{amssymb}
 \usepackage{amsmath}

\patchcmd{\theendnotes}
  {\makeatletter}
  {\makeatletter}
  {}{}

\newenvironment{sciabstract}{%
\begin{quote} \bf}
{\end{quote}}

\title{To vote, or not to vote? On the epidemiological impact of electoral campaigns at the time of COVID-19} 

\author
{Davide Cipullo,$^{1\ast}$ Marco Le Moglie$^{2}$\\
\\
\normalsize{$^{1}$Department of Economics, Uppsala University}\\
\normalsize{$^{2}$Department of Economics and Finance, Catholic University of the Sacred Heart}\\
\\
\normalsize{$^\ast$Correspondence to: davide.cipullo@nek.uu.se}
}

\date{}

\begin{document} 
\makeatletter
\makeatother

\baselineskip24pt

\maketitle 

\begin{sciabstract}

Elections are crucial for legitimating modern democracies, and giving all candidates the possibility to run a proper electoral campaign is necessary for elections' success in providing such legitimization. Yet, during a pandemic, the risk that electoral campaigns would enhance the spread of the disease exists and is substantive. In this work, we estimate the causal impact of electoral campaigns on the spread of COVID-19. Exploiting plausibly exogenous variation in the schedule of local elections across Italy, we show that the electoral campaign preceding this latter led to a significant worsening of the epidemiological situation related to the disease. Our results strongly highlight the importance of undertaking stringent measures along the entire electoral process to minimize its epidemiological consequences.

\end{sciabstract}

\newpage

\section{Introduction}
\label{intro}

Elections represent the primary tool for legitimating governments and their actions. This legitimization is crucial for the correct functioning of any democracy and becomes even more needed during a pandemic when decisions that limit individual rights substantially are taken. Yet, holding elections during a pandemic might enhance the spread of the disease due to different aspects of the electoral process, among which the candidates' campaign preceding the vote arguably represents the most problematic one. On the one hand, the possibility for all candidates to run a proper electoral campaign for correctly informing voters about their platform is necessary for the success of elections in legitimating the winner\cite{Lake2021}. On the other hand, the candidates' campaign induces a high frequency of in-person meetings where social distancing is challenging to enforce. 

In many countries, election schedules concur with the spread of the infection from SARS-CoV-2 \cite{WHO2020}, and the related illness known as COVID-19 \cite{IDEA2021}, thus imposing a trade-off between the accountability of elected officials and diffusion of the disease. Countries with elections planned in the year 2020 debated on whether to organize or postpone elections \cite{EU2020, IDEA2020b}. Where elections occurred, a wide range of country-specific precautionary policies was adopted to limit the contagion and ensure their smooth running \cite{IDEA2020}. Despite the importance of such a public debate, empirical evidence on the impact of electoral campaigns on the diffusion of COVID-19 is very scarce and incomplete \cite{Bernheim2020}. One prominent reason is that experimental variation is lacking: countries where elections did not occur, are not necessarily a reliable comparison group for countries where elections occurred \cite{Imbens2015}. 

In this work, we identify the causal impact of electoral campaigns on the diffusion of COVID-19 by exploiting a natural experiment taking place in Italy at the beginning of the \textit{second wave} of the pandemic. More specifically, on the same days as a national referendum to amend the constitution, regional elections took place in seven out of twenty regions. This plausibly exogenous variation to the intensity of the electoral campaign allows us to compare the epidemiological outcomes in areas with and without regional elections before and after the campaign's start.

 We find that the electoral campaign preceding the regional elections did lead to a significant increase in new infections (7\%), percentage of positive tests (15\%), ordinary hospitalizations (24\%), entries in intensive care units (5.3\%), and deaths (0.6\%) related to the disease. Taken together, our results inform the debate on how managing elections at the time of COVID-19 in two critical manners. First, they point towards the existence of a substantive impact of elections on the spread of COVID-19, thus proving the necessity to enforce a strict safety protocol to minimize such impact.  Second, they show that a large portion of the epidemiological risk connected to elections is concentrated during the electoral campaign preceding the vote, when sanitary precautions are more difficult to enforce than at the polling station.

\section{Background} \label{sec:background}

Since the first documented case of local transmission on February 20$^{\text{th}}$, 2020, the ascending number of infected individuals and that of hospitalizations and deaths related to COVID-19 induced the Italian government to enforce a strict lockdown over the entire national territory. The lockdown lasted more than two months (i.e., from March 11$^{\text{th}}$ to May $17^{\text{th}}$). By the end of May, the number of new daily cases stabilized around a few hundred. This level remained almost unaltered until the beginning of August when the number of new infections started to rise again.  
By the beginning of October, Italy was again experiencing a substantive increase in the daily number of new infections that led to a new set of public interventions to slow down the spread. Figure S.1 in the Supplementary Material shows a graphical overview of the evolution of daily infections, deaths, and hospitalized individuals.

On September 20$^{\text{st}}$ and 21$^{\text{st}}$,  at the dawn of the \textit{second wave} of the pandemic, Italy went to the polls. All Italians were asked in a referendum whether they intended to confirm a constitutional reform to reduce the number of members of the national parliament. The referendum was initially scheduled on March 29$^{\text{th}}$ and then postponed due to the epidemic. 

\begin{figure}[h!]
\centering
\caption{The Elections of Mid-September}\label{fig:campaign_intensity}
\subcaptionbox*{} {\centering\includegraphics[width=0.59\textwidth]{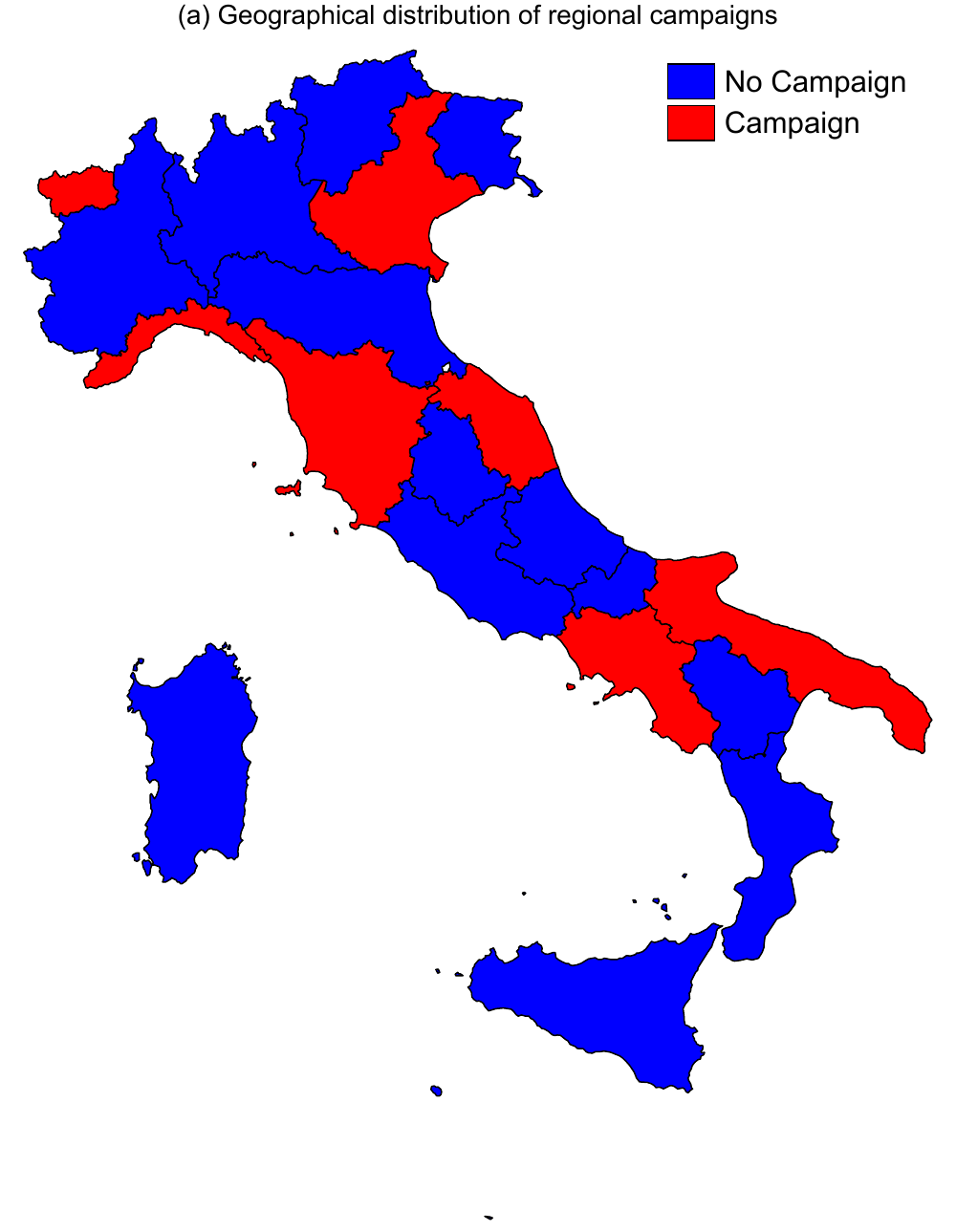}}
\subcaptionbox*{} {\centering\includegraphics[width=0.39\textwidth]{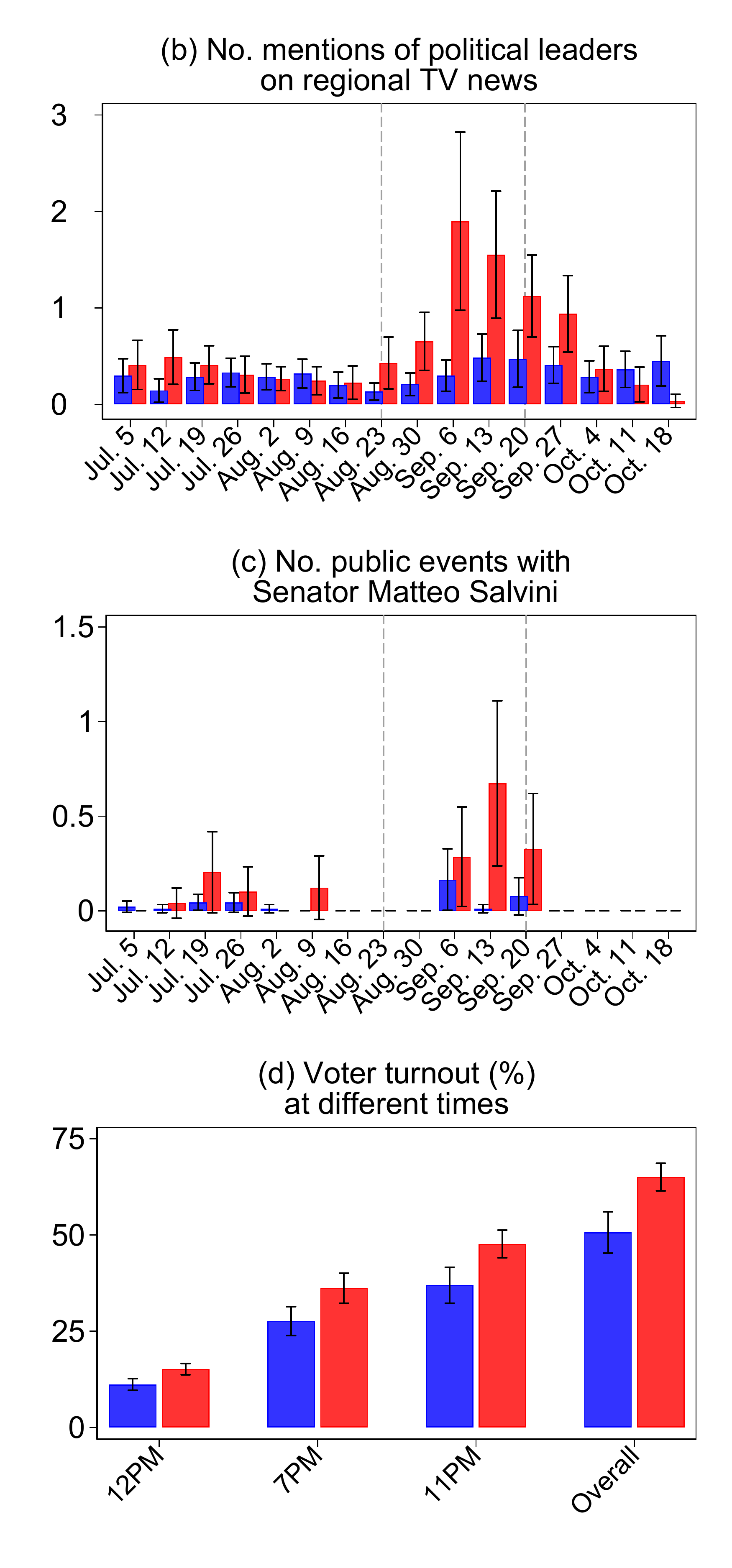}}
\caption*{\scriptsize{\textbf{Notes}: Red-colored areas and bars refer to regions voting for both the referendum and the renewal of their president and council, while blue-colored areas and bars refer to regions voting only for the referendum. Political leaders included in panel (b) are Di Maio, Meloni, Renzi, Salvini, and Zingaretti. Turnout measures in panel (d) are computed on September 20$^{\text{th}}$ at 12pm, 7pm, 11 pm and September 21st at 3pm and refer to participation in the referendum. In panels (b)-(d), bars represent sample averages by treatment status while vertical short line over each bar represents 95\% confidence intervals. In panels (b) and (c), the vertical dashed lines represent the date in which the electoral campaign officially started (August 21$^{\text{st}}$, lines on the left) and the election date (September 21$^{\text{st}}$, lines on the right).}}
\end{figure}

In the same days, seven out of twenty regions also went to the polls for renewing their regional president and council because of the regular expiration of the legislative term, according to an electoral calendar strictly disciplined by the law \cite{Regions}. Panel (a) of Figure \ref{fig:campaign_intensity} shows that such regions' geographical distribution is homogeneous across the country. Similar to the referendum, regional elections were postponed from their original schedule dates in the Spring. 

Regardless of the type of voting held in each region, the national government implemented a strict protocol to guarantee safety at the polling stations. This protocol included physical distancing between different voting booths, compulsory masks for all voters and workers, and the use of hand sanitizers before touching the ballot or the voting pencil. 

The referendum did not have to achieve any participation quorum to be valid, thus ultimately narrowing citizens' incentives to go to the polling station. Moreover, since the constitutional reform underneath the referendum had the support of most parties represented in the parliament, the political campaign carried on in favor and against the confirmation was almost absent, or at least very mild \cite{Reform}. 

Regional elections, instead, were an important battleground. First, many candidates competed for a regional council seat, which is a powerful and well-rewarded position. Second, regions are assigned by the constitution the authority on healthcare policies -- which is a vital issue at stake during a pandemic. Third, the opposition motivated their voters to signal their dissent to the national government, and, by contrast, members of the government coalition campaigned in support of their candidates. As a result, candidates and party leaders run an intense campaign both on the media and in-person. For instance, panel (b) of Figure \ref{fig:campaign_intensity} reports the average number of national political leaders' mentions on regional TV news. The figure shows that, in the period between the official start of the electoral campaign and the election days, such number was significantly higher in regions voting for renewing their government than in the others. Moreover, panel (c) of Figure \ref{fig:campaign_intensity} shows that the opposition leader sen. \textit{Matteo Salvini} (\textit{Lega}) participated in the same period to a much greater number of public events in the former regions. In turn, as shown in panel (d) of Figure \ref{fig:campaign_intensity}, voter turnout was approximately 20 percentage points higher in regions where both the referendum and the electoral round took place than in those voting only for the former \cite{turnout}. 
\section{Data} \label{sec:data}

We rely on the daily epidemiological bulletin on the spread of COVID-19 in Italy, compiled by the Department of Civil Protection (DPC), the governmental agency responsible for monitoring the disease's evolution across the country. From the information contained in the bulletin, we calculate six epidemiological outcomes measuring the spread of COVID-19, all normalized per 100,000 inhabitants. Specifically, we focus on i) the daily number of new infections, ii) the daily number of tests performed, iii) the daily percentage of positive tests, iv) the daily number of individuals currently undertaking ordinary care in hospitals, v) the daily number of patients currently in intensive care units (ICUs), and vi) the daily number of deaths. 

To interpret coefficients as percentage changes, we perform a logarithmic transformation of all outcomes. To limit the noise due to delay in reporting, we use the seven-day moving average within each region. We focus on bulletins published between June 29$^{\text{th}}$ and October 15$^{\text{th}}$ to keep the comparison across regions as reliable as possible. Indeed, the first day corresponds to the day recording the minimum number of cases between the first and the second Italian wave of the pandemic at the national level. In contrast, regions started to diverge in their adoption of specific restrictions to limit the second wave on the last day included in the sample.

To ensure that different regions were experiencing similar risks of spread in the period under consideration, we also employ regional data from the \textit{Google COVID-19 Community Mobility Reports}. This data report daily changes in individual mobility among workplaces, retailing stores, stations, and residential areas compared to the beginning of the epidemic. 

In the Supplementary Material, we assess the robustness of our findings by employing information on the number of municipal elections held in each region on September 20-21$^{\text{st}}$ as well as several geographic, socio-economic, and demographic characteristics from the Italian National Institute of Statistics (ISTAT). Appendix S.2 in the Supplementary Material provides a detailed list and description of all variables employed in our analysis, while Table S.1, still in the Supplementary Material, shows the summary statistics for all of them.

\section{The electoral campaign and the spread of COVID-19} \label{sec:results}

We compare, in a Difference-in-Differences setting (DiD), regions that held both the referendum and the regional election on September 20-21$^{\text{st}}$ -- the treatment group -- with regions only voting for the referendum -- the control group -- before and after the official start of the electoral campaign -- 30 days before the vote. 

Treatment assignment is plausibly exogenous because different regions vote in different years depending on their electoral cycle timing that usually repeats every five years and law disciplines calendars of voting and campaign dates. Moreover, the national government announced the new election calendar in the mid of July when the diffusion of COVID-19 was at its minimum level.

In particular we estimate the following linear regression model 
\begin{equation}
\label{eq:did}
y_{r,t}=\gamma_0+\gamma_1 Campaign_{r} \times Post_{t} +  \zeta^{'} \textbf{X}_{r,t} + \epsilon_{r,t} , \,
\end{equation}

where $y_{r,t}$ is one of the epidemiological outcomes measured in the region $r$ at date $t$, $Campaign_r$ is a dummy equal to one for treatment regions, $Post_t$ is a dummy equal to one for all dates after the official starting of the electoral campaign. $\textbf{X}_{r,t}$ is a set of control variables which include: i) region fixed effects aimed at capturing all the time-invariant heterogeneity across regions; ii) regional-by-week linear pre-treatment trends allowing to control for the linear effect of weekly shocks preceding the treatment and that are specific to each region; iii) adjacent regions-date fixed effects that control for daily shocks common to adjacent treatment and control regions. 
We account for serial and spatial correlations in the error terms $\epsilon_{r,t}$ using Conley-HAC standard errors \cite{conley1999} with maximum cutoffs for both spatial and time lags.

\begin{figure}[h!] 
\centering
\caption{The Dynamics of Individual Mobility Through Google's Data}\label{fig:google}
 
\subcaptionbox{: Workplaces} {\includegraphics[width=0.49\textwidth]{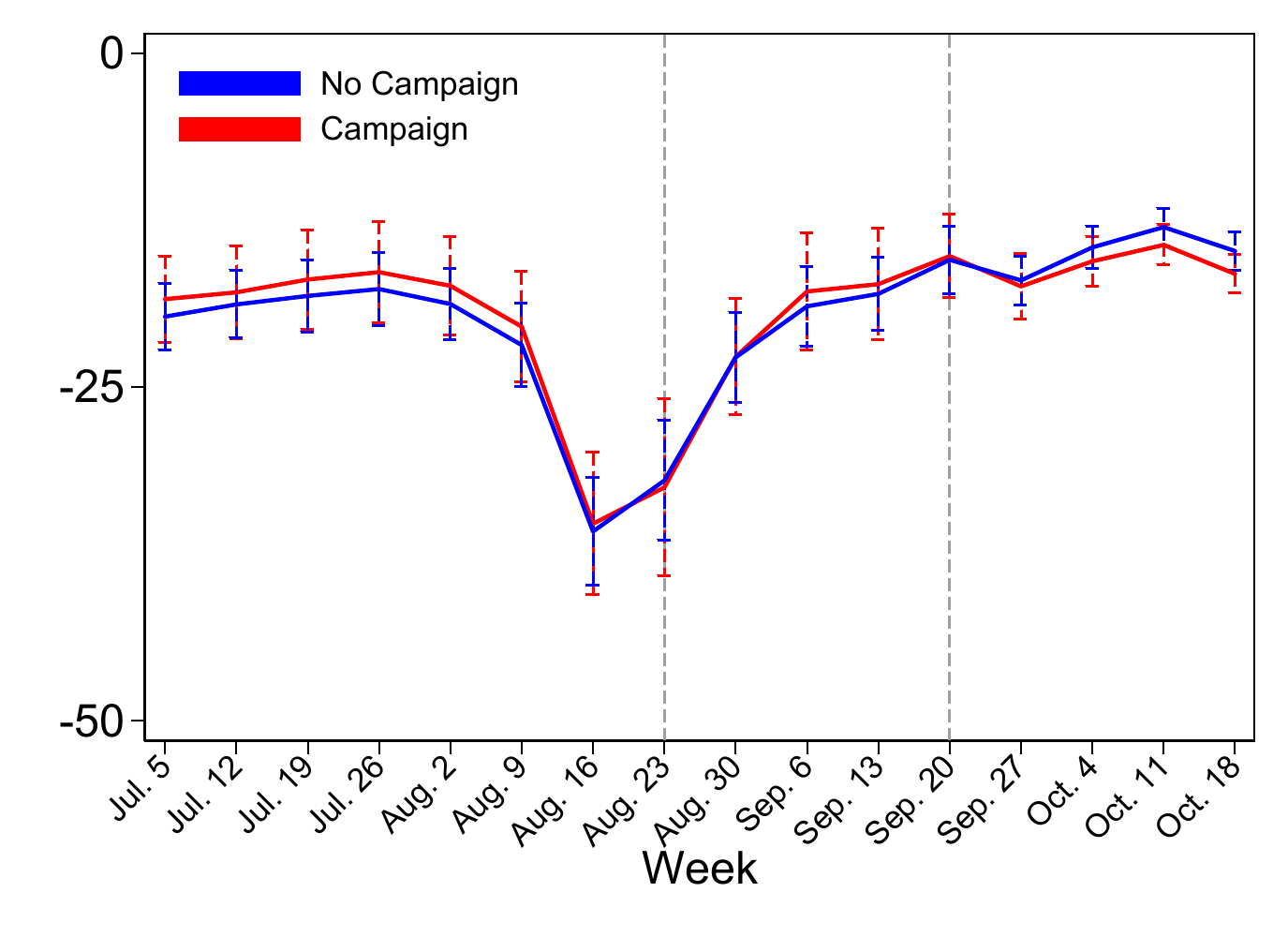}}
\subcaptionbox{: Retailing} {\includegraphics[width=0.49\textwidth]{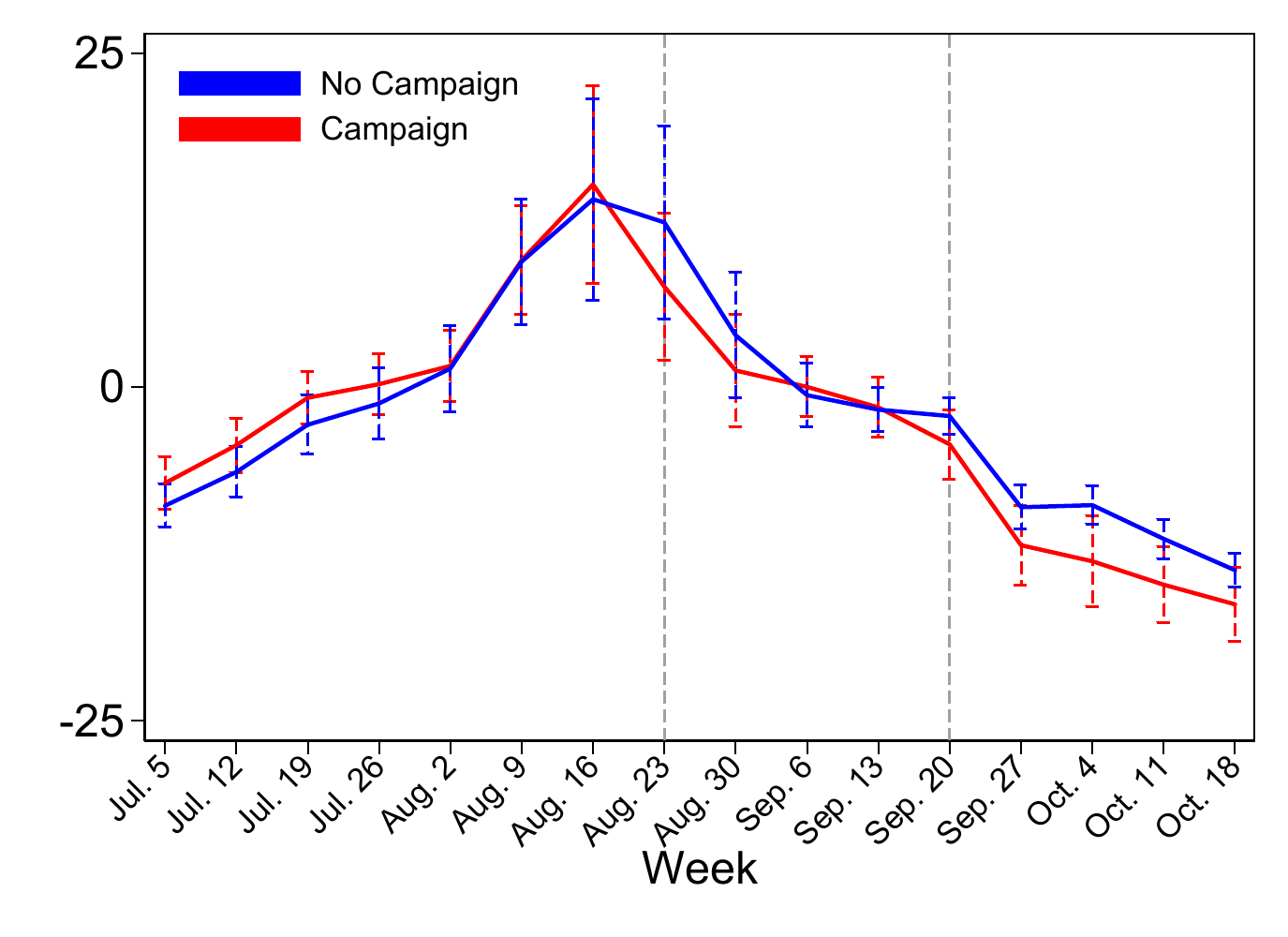}}

\vspace{.01cm}
\subcaptionbox{: Stations} {\includegraphics[width=0.49\textwidth]{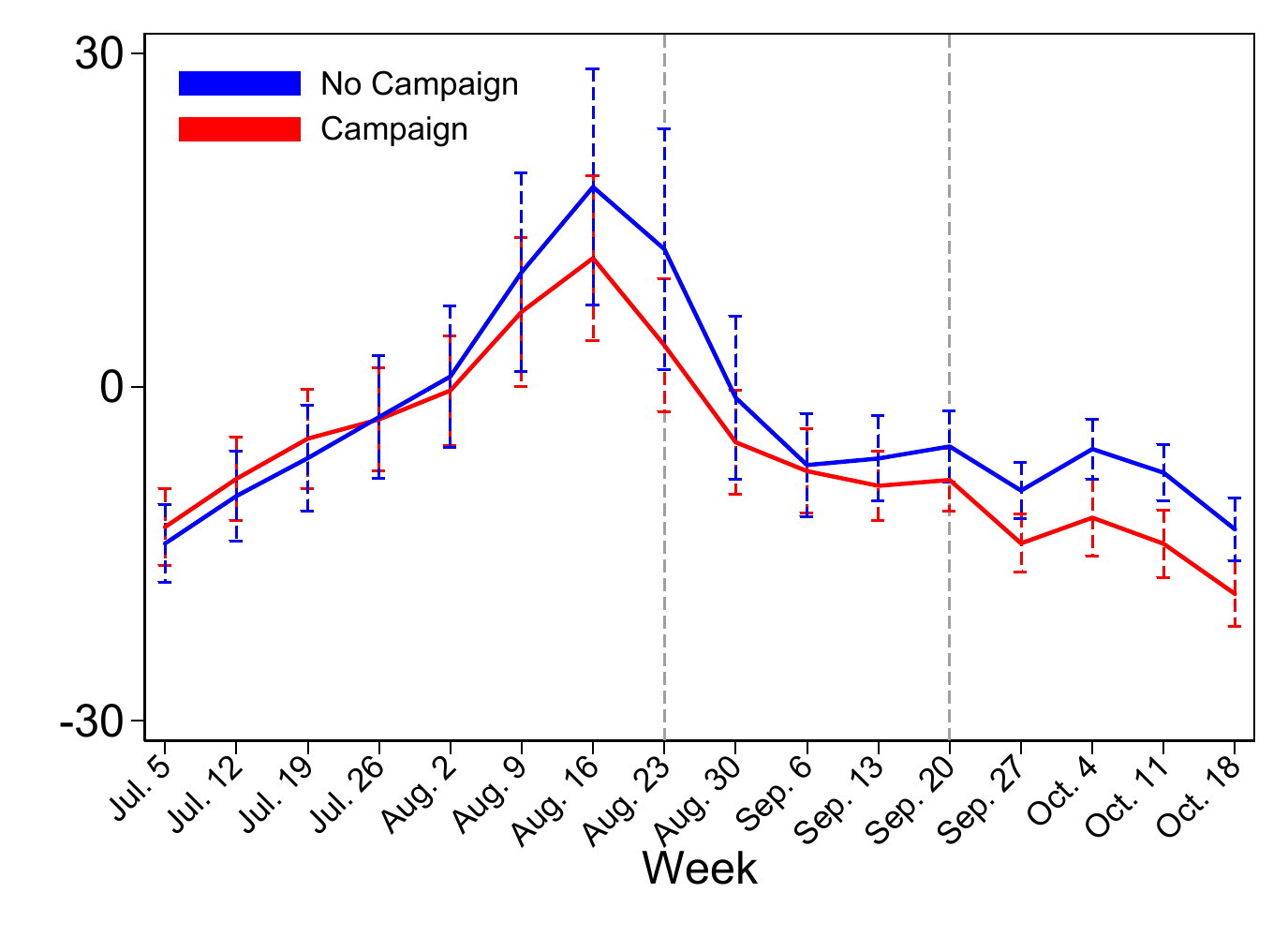}} 
\subcaptionbox{: Residential} {\includegraphics[width=0.49\textwidth]{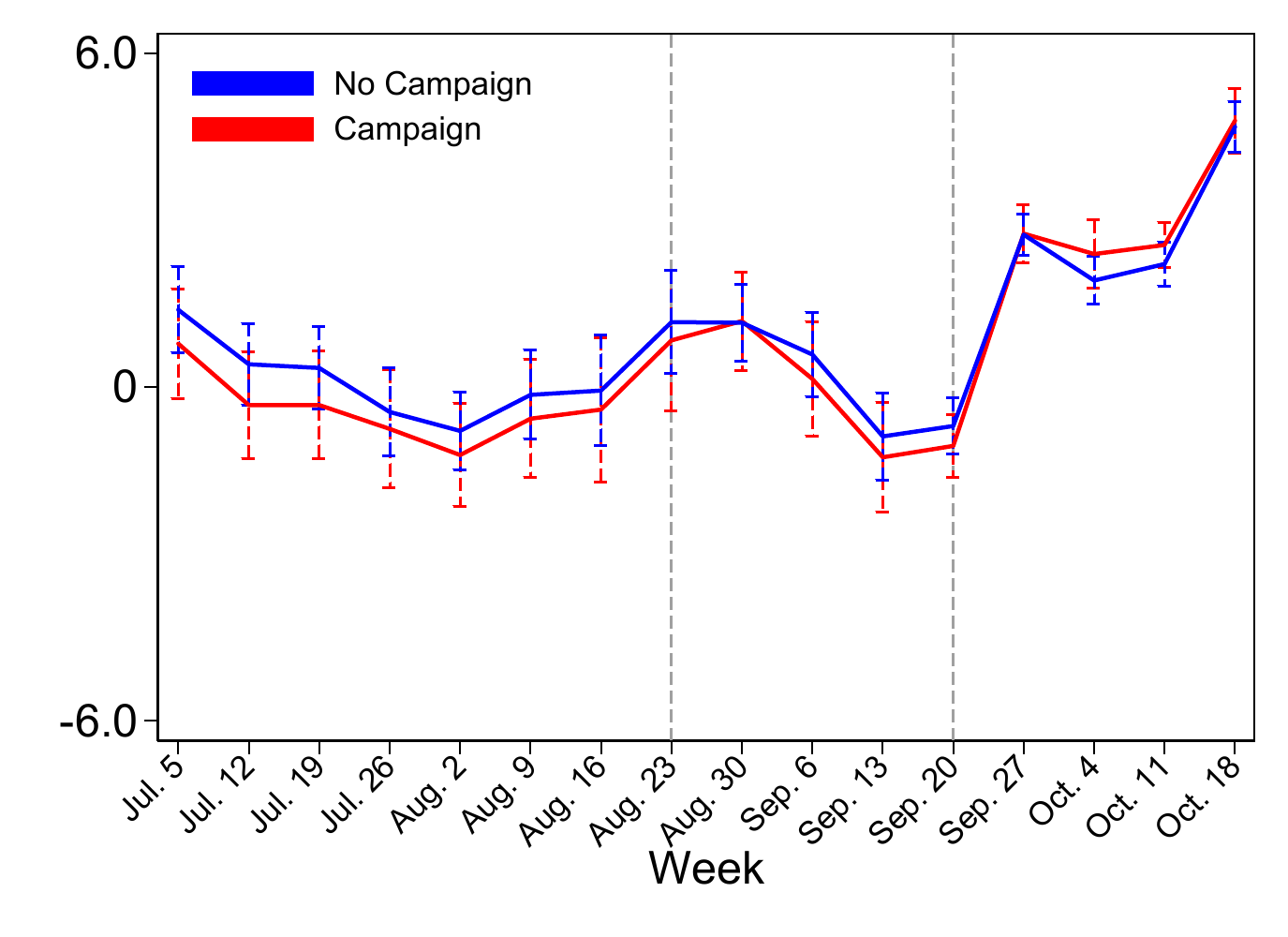}} 
\caption*{\scriptsize{\textbf{Notes}: The figures report the average and 95\% confidence intervals of the daily changes in individual mobility compared to February 2020, in treatment (red line) and control (blue line) regions. In all panels, the vertical dashed lines represent the date in which the electoral campaign officially started (i.e. August 21st, lines on the left) and the week in which the election took place (i.e., September 20-21st, lines on the right).}}
\end{figure}

The coefficient of interest is $\gamma_1$, representing the differential effect of the electoral campaign between treatment and control regions. Yet, its causal interpretation relies on the actual comparability between the two groups of regions, particularly on the assumption that, in the absence of the electoral campaign, the dynamics of the outcomes would have been the same between the two groups. Our analysis deals with the validity of this assumption in different manners. First, in Figure \ref{fig:google} we exploit \textit{Google}'s data on individual mobility to show the absence of significant differences in mobility patterns between treated and control regions over the entire observational window. This implies that the two groups of regions were indeed experiencing similar risks of spread in the period under consideration. Second, by including in our regression adjacent regions-date fixed effects, we are constantly comparing changes in the outcome of interest between regions that share part of their border and are thus more comparable. Third, the inclusion of adjacent region-date fixed effects also helps to rule out the impact of weekly trends among adjacent regions, such as those related to summer holidays. Fourth, the inclusion of regional-by-week linear pre-treatment trends should capture several potential region-specific differences in the dynamics of the outcomes before the electoral campaign, making treatment and control regions strongly comparable once the campaign starts.

Lastly, we propose a dynamic specification in which the dummy for the post-treatment period ($Post_t$) is replaced with a set of week dummies to empirically exclude that the dynamics of the six epidemiological outcomes were significantly different between treatment and control regions also before the campaign. This specification also allows us to get further insights on the dynamics of the effect after the beginning of the campaign. More specifically, we estimate the following regression model: 
\begin{equation}
\label{eq:event}
y_{r,t} =  \gamma_0 + \underset{w\in{[-6, 8], w\neq{-1}}}{\sum}\gamma_{w}{Campaign_r}\times \textbf{1}(Week_t=w) +  \zeta^{'} \textbf{X}_{ij} +\epsilon_{r,t} \,,
\end{equation} 

Table \ref{tab:table_results_region} presents the estimates for Equation \ref{eq:did}. The daily number of new infections rose by about 7\% (p-value=0.037) due to the electoral campaign. Even if significantly different from zero, this coefficient probably reflects a lower bound of the electoral campaign's impact on the number of new infections, given that the campaign also led to a reduction in the daily number of tests performed equal to 8.3\% (p-value$<$0.001). Since the Italian health system is under the authority of the regional governments, the reduction in the number of tests might reflect a simple increase in uncertainty among regional bureaucracies induced by the electoral process, a strategical behavior by incumbent governments to secure their re-election, or rather a combination of the two. Regardless of its actual origins, once we account for the reduction in testing, the electoral campaign's epidemiological impact appears to be sizeable. In particular, the percentage of positive COVID-19 tests rose by 15\% (p-value$<$0.001). The substantive impact further reflects into a significant increase in the daily number of ordinary hospitalizations, the number of people hospitalized in ICUs, and deaths related to the epidemic,  which since the start of the former rose by 24\% (p-value$<$0.001), 5.3\% (p-value=0.007) and 0.6\% (p-value=0.011), respectively.

\begin{figure}[h!] 
\centering
\caption{The Impact of Electoral Campaign on the Spread of COVID-19: Dynamic Specification}\label{fig:figure_trends_region}
\subcaptionbox{: New infections} {\includegraphics[width=0.35\textwidth]{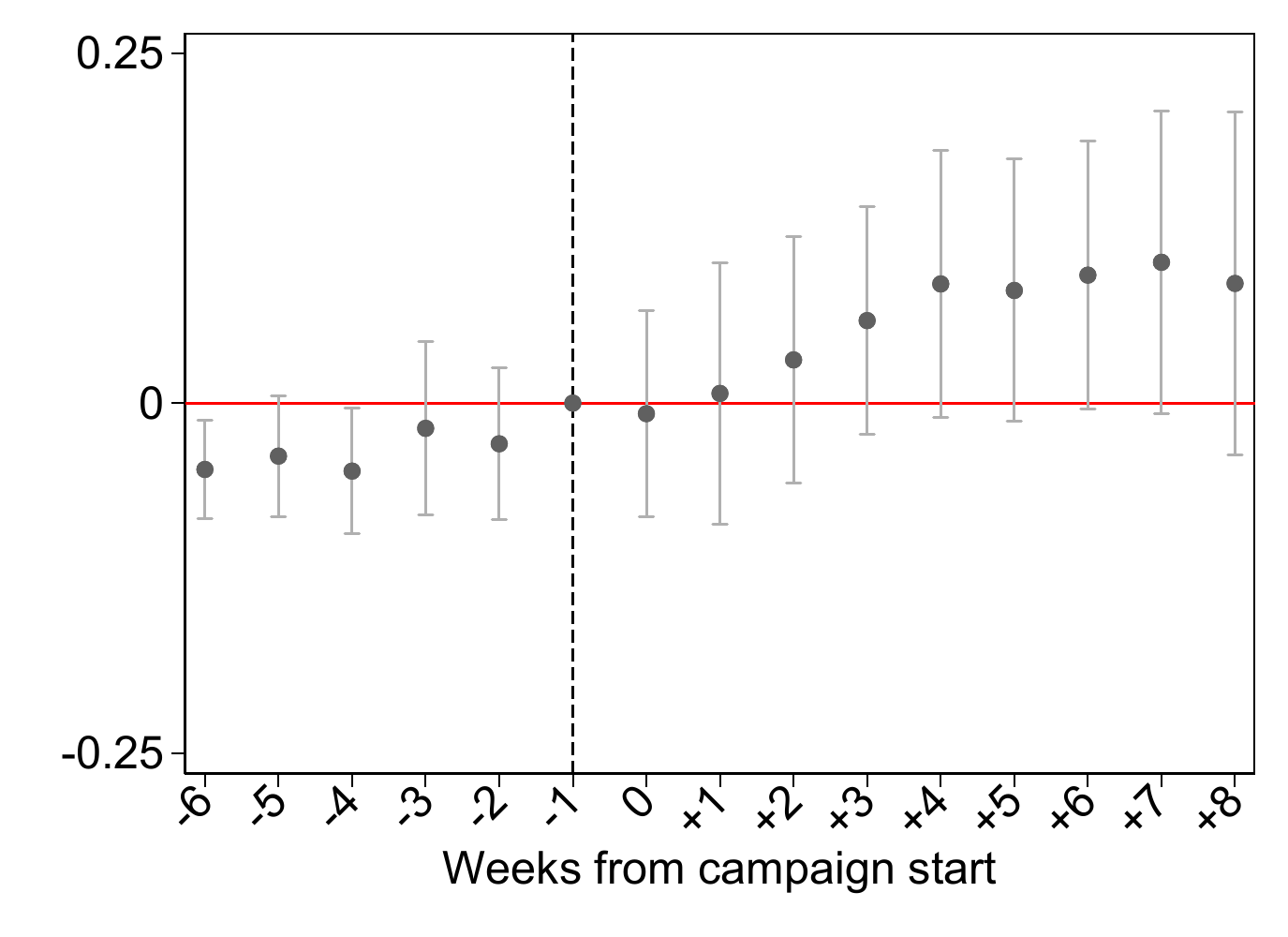}} 
\hspace{1cm}
\subcaptionbox{: Tests} {\includegraphics[width=0.35\textwidth]{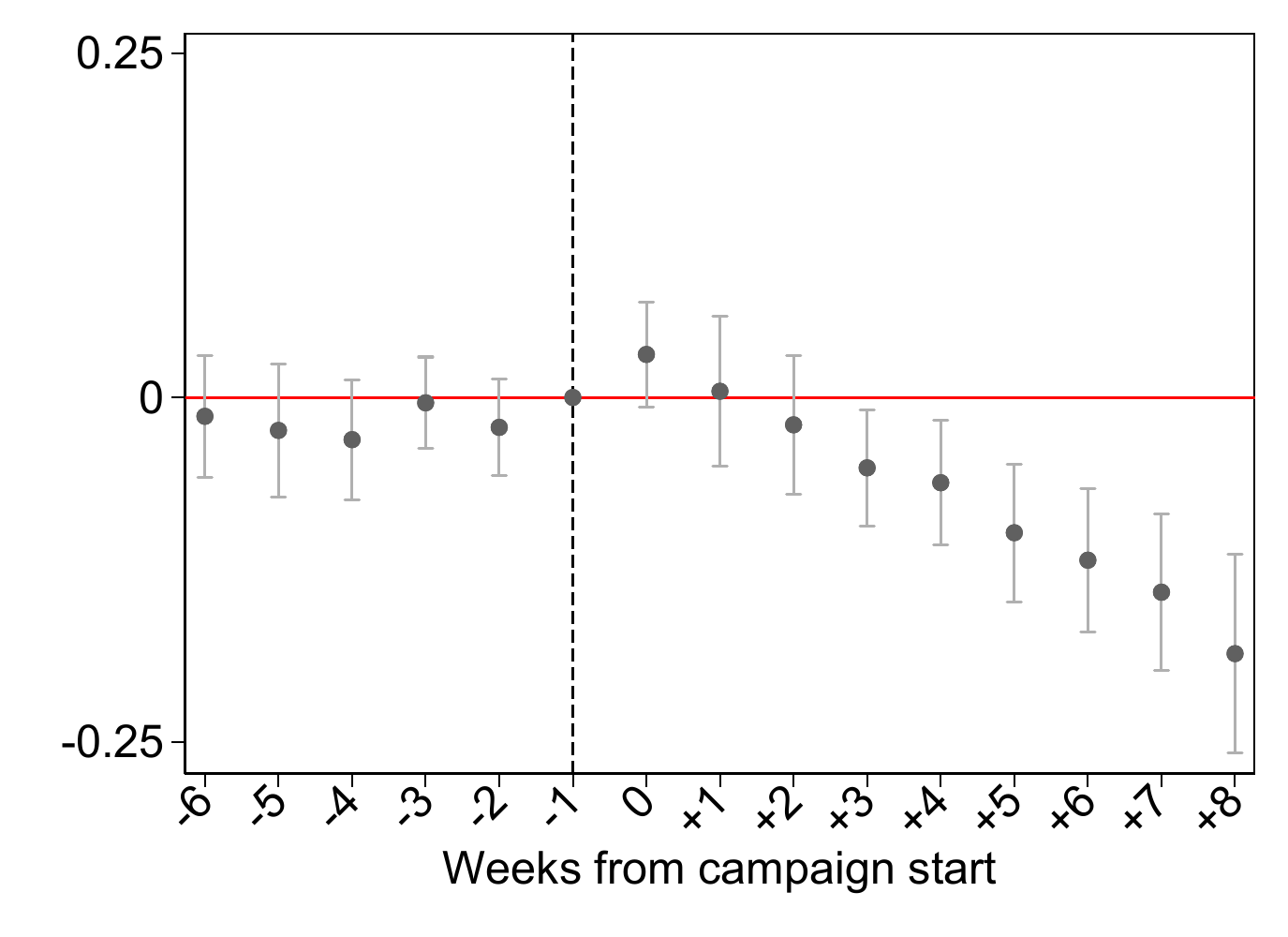}} 

 \vspace{.01cm}

\subcaptionbox{: \% Positives} {\includegraphics[width=0.35\textwidth]{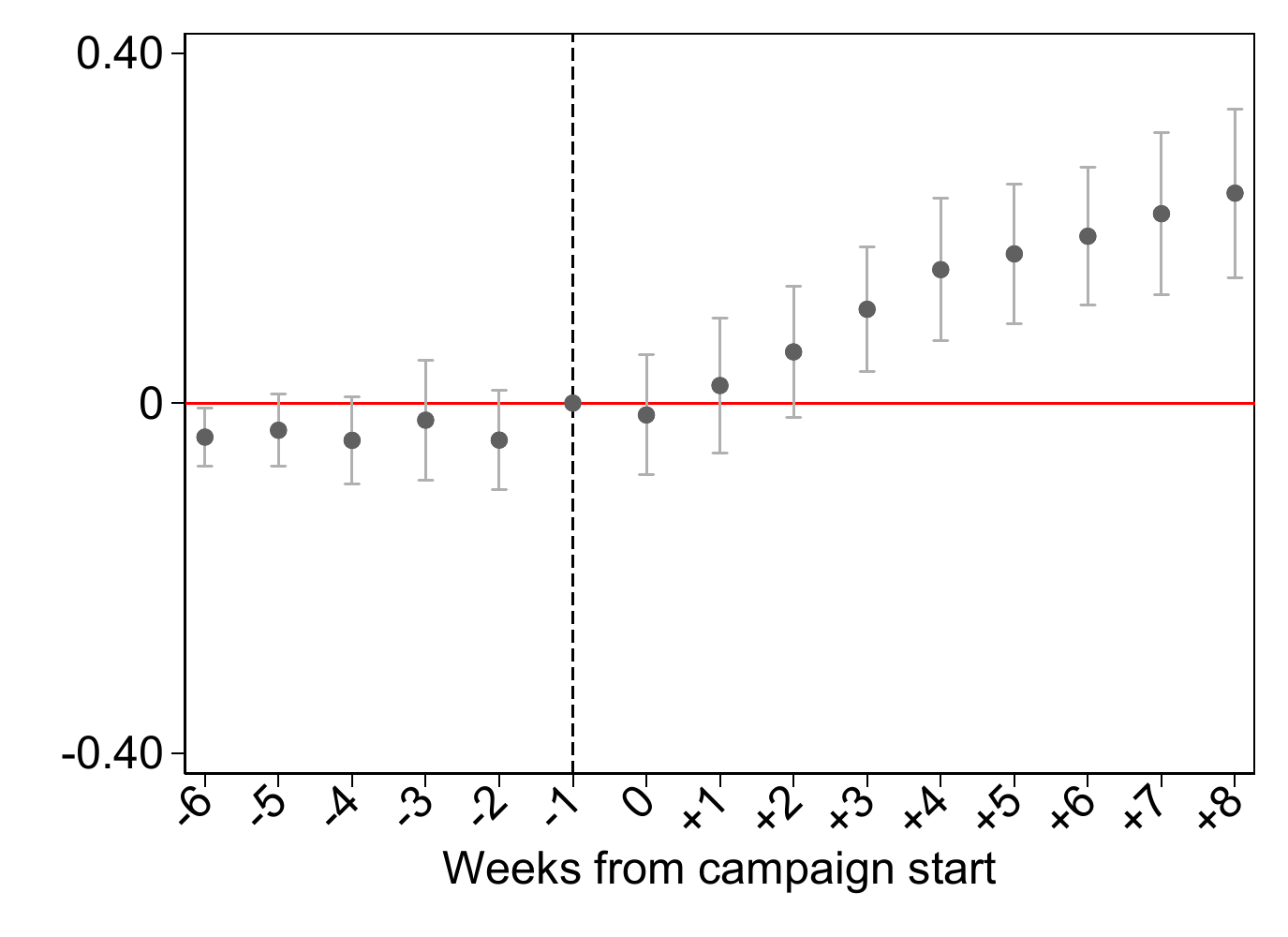}}
\hspace{1cm}
\subcaptionbox{: Hospitalized} {\includegraphics[width=0.35\textwidth]{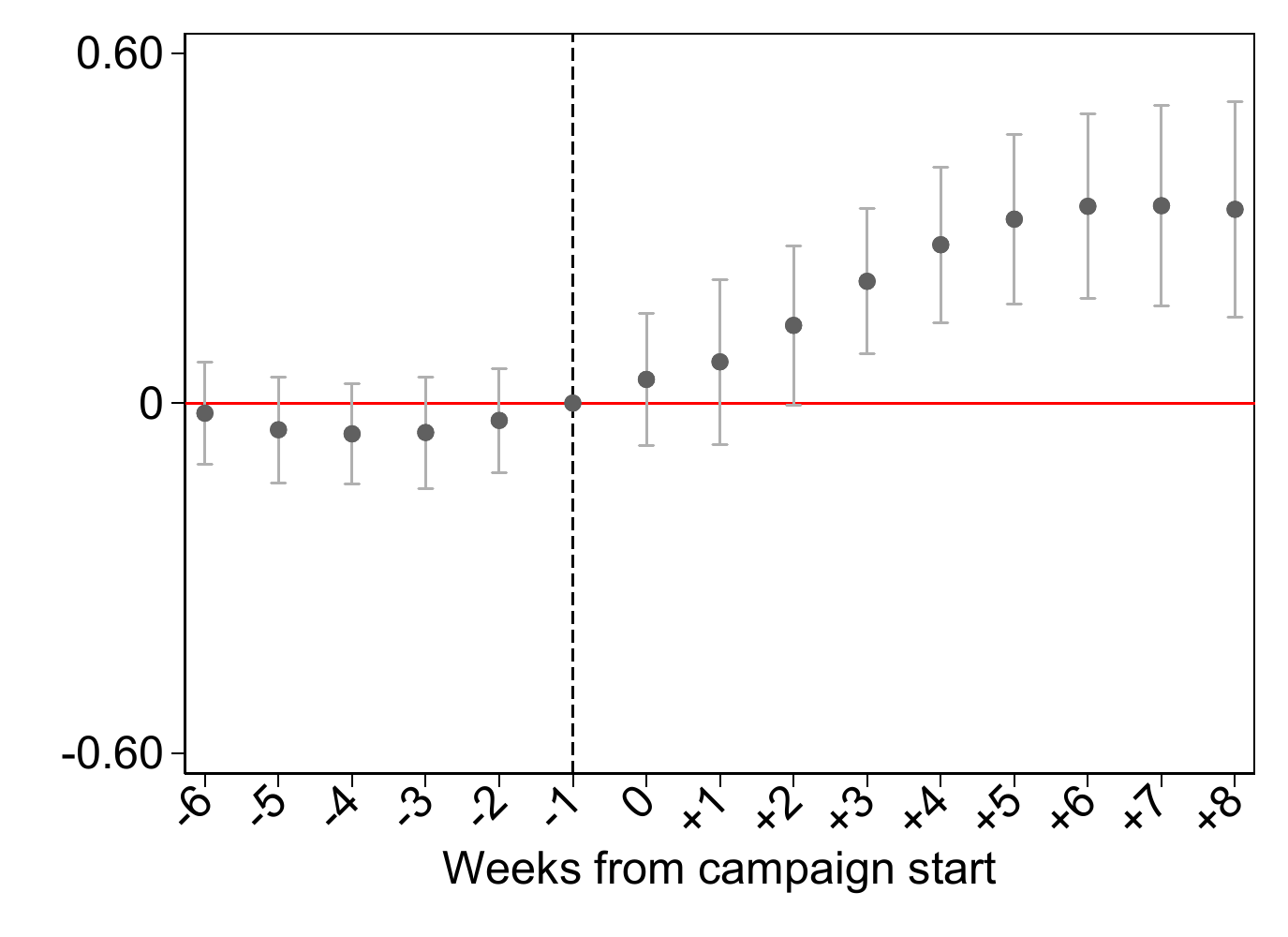}}

\vspace{.01cm}

\subcaptionbox{: ICUs} {\includegraphics[width=0.35\textwidth]{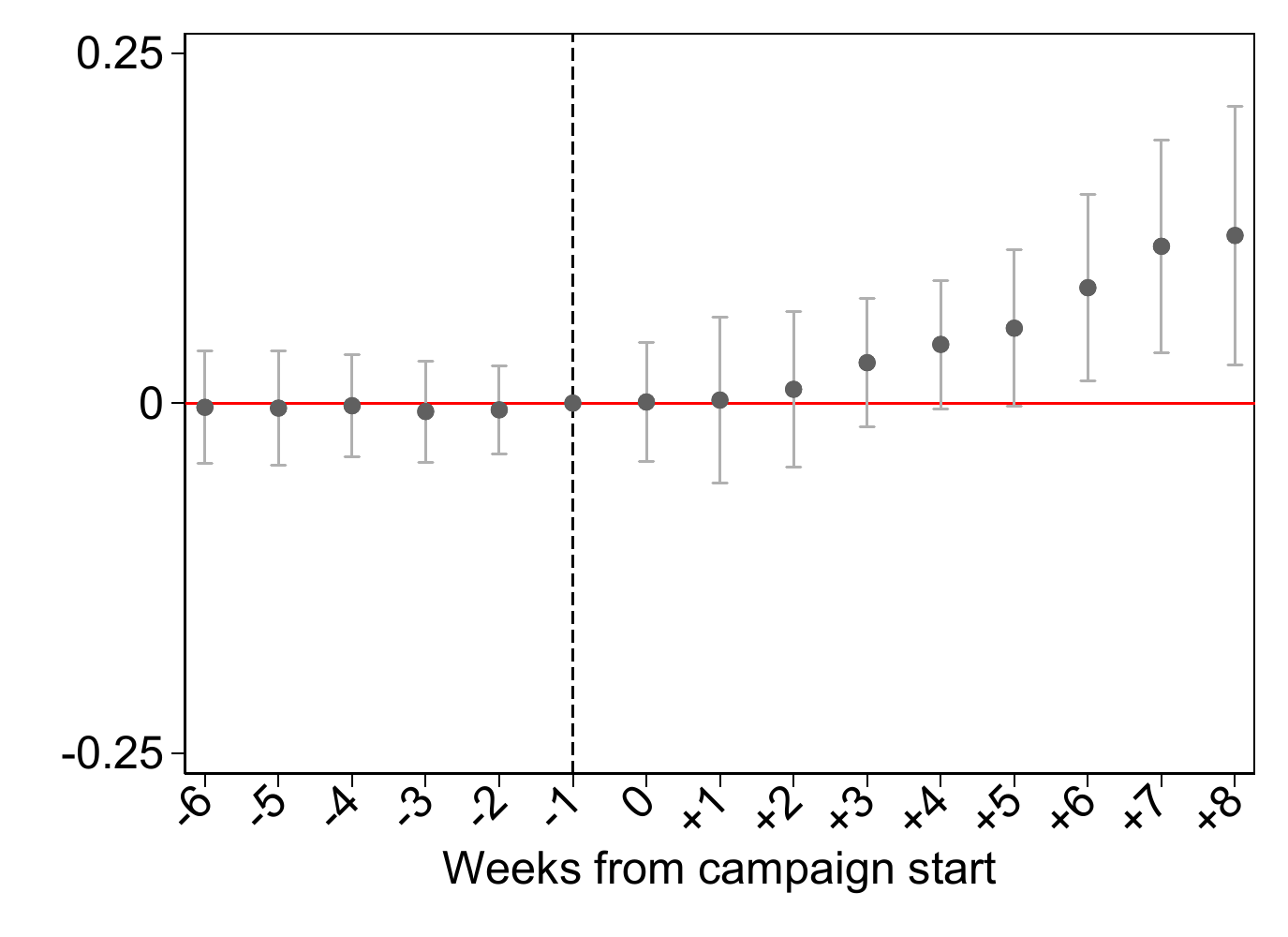}} 
\hspace{1cm}
\subcaptionbox{: Deaths} {\includegraphics[width=0.35\textwidth]{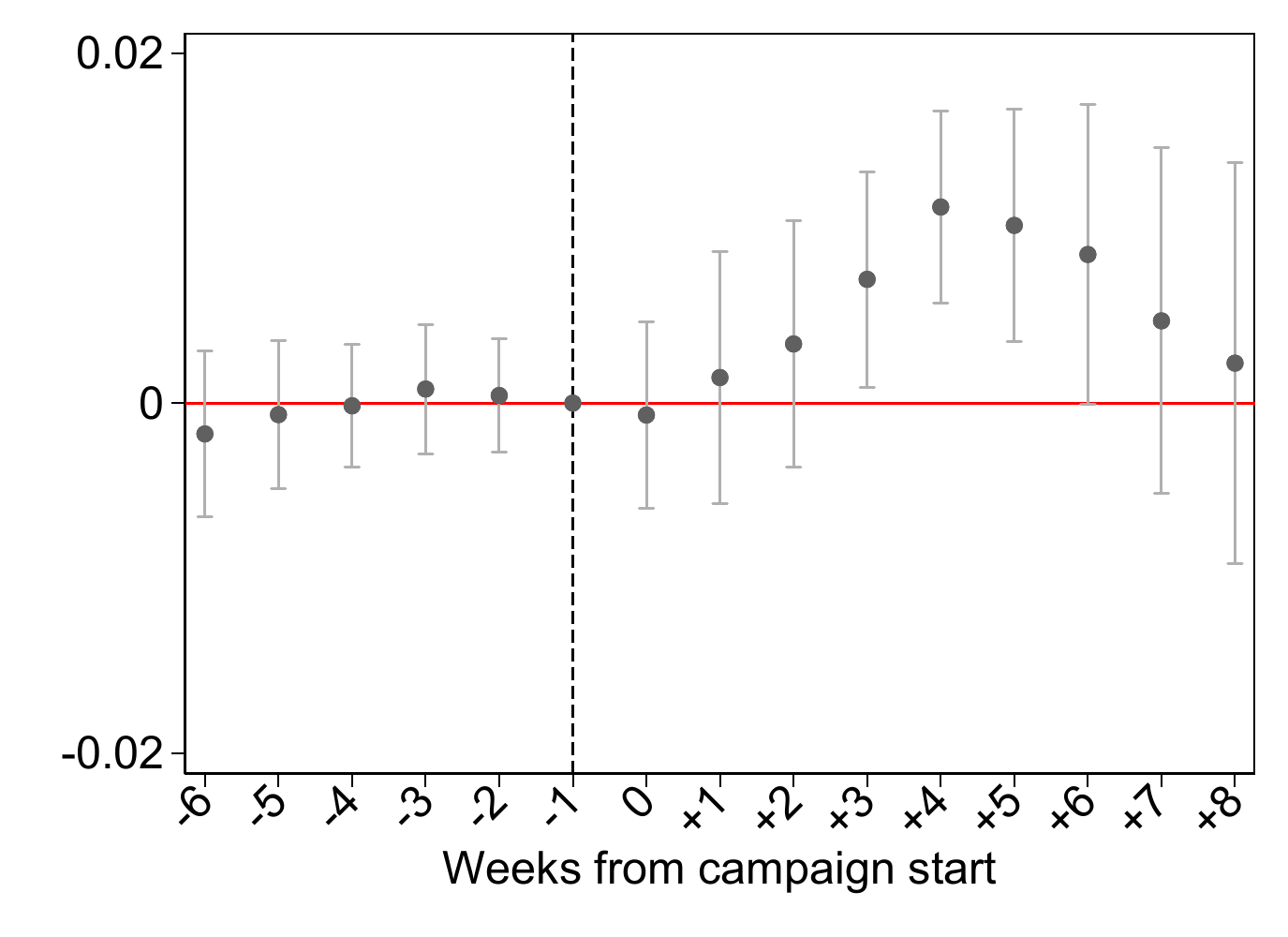}} 
\caption*{\scriptsize{\textbf{Notes}: The figures report estimates and 95\% confidence intervals for the coefficients of the interactions included in Equation \ref{eq:event}, between the dummy $Campaign$, which is equal to one for treatment regions, and the week dummies. $week=-1$ is the reference week (i.e., week 33 of 2020). All dependent variables are the natural logarithm of the seven-day moving average of the daily number at the regional level. See Section \ref{sec:data} for details. $Campaign$ is a dummy equal to 1 for treatment regions, and 0 otherwise.
Confidence intervals are based on standard errors that are robust to serial and spatial correlation and computed imposing maximum cutoff for both spatial and time lags. In all panels the vertical dashed line represents the date in which the electoral campaign officially started (i.e. August 21st).}}

\end{figure}

Figure \ref{fig:figure_trends_region} plots the estimated coefficients for the interaction terms of Equation \ref{eq:event}. First, we notice the absence of differential pre-treatment dynamics in all epidemiological outcomes between treatment and control regions before the campaign's official start. The absence of pre-treatment dynamics is crucial to supporting the actual comparability between the two groups of regions, and ultimately the possible causal interpretation of our results. Second, the treatment effect dynamics reveal both similarities and differences across outcomes. For all outcomes, treatment effects become visible with a certain lag from the beginning of the campaign and become statistically significant at the 5\% critical level only after some weeks. These patterns are consistent with a cumulative epidemiological effect of the electoral campaign and the 7-14 days' average incubation period of the disease \cite{Qin2020}. 

While most of the outcomes stabilize after the fourth post-treatment week, few others still increase, like the percentage of positive tests and the number of hospitalizations in ICUs. Concerning the former, its increase after the fourth week is entirely due to the further reduction in the testing capacity. Regarding the latter, hospitalizations in ICUs still increase to the sixth week after the campaign's start and then finally stabilized. The fourth post-treatment week corresponds to the week in which the campaign ends, and the vote occurs. Therefore, this study's results suggest that elections enhanced the spread of COVID-19 primarily through the campaign preceding the vote.

In the Supplementary Material, we assess the robustness of the estimates presented in Table \ref{tab:table_results_region} with respect to i) the inclusion of controls for linear pre-treatment trends in several demographic, social, and economic characteristics at the regional level (Table S.2); ii) changes in the starting week of the electoral campaign (Figure S.2); iii) the exclusion from our sample of one region at the time  (Figure S.3); iv) the use of different spatial and time cutoffs for calculating the Conley HAC standard errors  (Figure S.4). The estimates survive all these tests, reassuring us about the empirical design's validity and the causal interpretation of the findings.

\section{Conclusions}\label{sec:conclusions}
In this study, we have estimated the causal impact of electoral campaigns on the spread of COVID-19 in Italy. By exploiting variation in the elections schedule across Italian regions in a Difference-in-Differences setting, we have found that the electoral campaign increased the spread of COVID-19 significantly. More specifically, our results show that the incidence of the disease, the number of patients seeking care in hospitals, and the number of deceased individuals increased more steeply in the treated than in the control regions at the dawn of the pandemic's \textit{second wave} in Italy. 

Our study's findings lead to a number of concluding remarks and policy recommendations. First, our findings show that the electoral campaign also negatively impacted testing capacity, which in Italy is under the authority of regional governments. Our study is agnostic on whether such reduction was due to the temporary uncertainty or intentional. However, reduced testing capacity can contribute to the spread of the disease because infected asymptomatic individuals are not isolated promptly, and their close contacts are not traced \cite{Ferrettieabb6936, Fetzer2020}. Reduced testing capacity eventually represents an additional channel through which electoral campaigns can affect the spread of the disease, and thus to eventually consider for minimizing their epidemiological impact.

Second, the comparison of the estimates' magnitudes suggests that the electoral campaign contributed to the spread of the disease more in low-risk groups than among high-risk individuals. In particular, the number of hospitalized patients in ordinary units increased by 24\%, while the number of individuals in ICUs and deceased increased by 5.3\% and 0.6\%, respectively. One possible interpretation is that individuals belonging to risk groups are less likely to spend time outside their home location \cite{Monodeabe8372} and participate in large meetings.

Finally, our findings shed light on the negative epidemiological impact of gatherings during which the enforcement of prevention policies is limited. Authorities should bear the epidemiological costs of electoral campaigns when deciding whether to organize elections and deciding which measures to take. While several prevention measures can be successfully implemented to limit the spread of the disease at the polling station -- distancing, face masks, hand sanitizers, and postal voting -- it remains an open question of how to conduct electoral campaigns safely.

\clearpage

\clearpage

\begin{table}[htbp]
\centering
\resizebox{1\linewidth}{!}{
\begin{threeparttable}
\normalsize
\caption{The Impact of the Electoral Campaign on the Spread of COVID-19}\label{tab:table_results_region}
\begin{tabular}{lcccccc} \hline\hline
  & (1) & (2) & (3) & (4) &(5) & (6)   \\
& Infections & Tests &  \% Posit.  & Hospit.  &  ICUs & Deaths  \\ 
 \hline
 &  &  &  & &  &   \\
 Campaign $\times$ Post & 0.070*&-0.083***&0.146***&0.240***&0.053**&0.006* \\
 & (0.033)&(0.019)&(0.031)&(0.052)&(0.020)&(0.002)\\
 &  &  &  &   &  &    \\
Observations &2,060&2,060&2,060&2,060&2,060&2,060\\
Baseline (value) &0.315&83.578&0.404&0.904&0.0032&0.012\\
Region  FE &\checkmark&\checkmark&\checkmark&\checkmark &\checkmark&\checkmark  \\
Regional pre-trends & \checkmark&\checkmark&\checkmark&\checkmark&\checkmark&\checkmark \\
 Date $\times$ Adjacent reg. FE &\checkmark&\checkmark&\checkmark&\checkmark&\checkmark&\checkmark\\
 \hline\hline
\end{tabular}

\caption*{\scriptsize{Notes: The table shows the differential impact of electoral campaign between regions voting both for the referendum and the regional election (treatment) and those only voting for the referendum (control), as obtained through the estimation of Equation \ref{eq:did}.
All dependent variables are the natural logarithm of the seven-day moving average of the daily number at the regional level. See Section \ref{sec:data} for details. $Campaign$ is a dummy equal to 1 for treatment regions, and 0 otherwise. $Post$ is a dummy equal to 1 for all the dates since August 21st, and 0 otherwise. Standard errors are robust to serial and spatial correlation and computed imposing maximum cutoff for both spatial and time lags. $\dagger$,  *,**,*** represent the 10\%, 5\%, 1\%, 0.1\% significance levels, respectively.}}
\end{threeparttable}
}
\end{table}

\clearpage

\clearpage \newpage

\section*{References and Notes:}
\nocite{*} 
\bibliographystyle{Science}

\begin{thebibliography}{10}
\bibitem{Lake2021}
J.~Lake, J.~Nie, Did Covid-19 Cost Trump the Election? \textit{CESifo Working Paper No. 8856}, (2021).


\bibitem{WHO2020}
World Health Organization, Modes of transmission of virus causing COVID-19: implications for IPC precaution recommendations. \textit{Scientific brief} 27 March 2020, (2020).


\bibitem{IDEA2021}
International Institute for Democracy and Electoral assistance (IDEA), Global overview of COVID-19: Impact on elections, available at \href{https://www.idea.int/news-media/multimedia-reports/global-overview-covid-19-impact-elections}{International IDEA's web page}, (2021). 


\bibitem{IDEA2020b}
International Institute for Democracy and Electoral Assistance (IDEA), Elections and COVID-19, International IDEA Technical Paper 1/2020, (2020).

\bibitem{EU2020}
A.~Radjenovic, R.~Mańko, G.~Eckert, Coronavirus and elections in selected Member States. \textit{European Parliamentary Research Service Briefings} PE651.969, (2020). 

\bibitem{IDEA2020}
International Institute for Democracy and Electoral assistance (IDEA), Risk mitigation measures for national elections during the COVID-19 crisis, available at \href{https://www.idea.int/news-media/news/risk-mitigation-measures-national-elections-during-covid-19-crisis}{International IDEA's web page}, (2020). 


\bibitem{Bernheim2020}
B.D.~Bernheim, N.~Buchmann, Z.~Freitas-Groff, S.~Otero, The Effects of Large Group Meetings on the Spread of COVID-19: The Case of Trump Rallies. \textit{Stanford, Institute for Economic Policy Research (SIEPR) Working Papers} 20(43), (2020). 














\bibitem{Imbens2015}
G.~Imbens, D.~Rubin, Causal Inference for Statistics, Social, and Biomedical Sciences: An Introduction. Cambridge: Cambridge University Press, (2015).


\bibitem{Regions}
These regions are Valle d'Aosta, Veneto, Liguria, Toscana, Marche, Campania, and Puglia.

\bibitem{Reform}
The reform was approved by the Italian parliament in October 2019, with 97\% of deputies and 80\% of senators in its favor, and confirmed by 70\% of the voters at the referendum.


\bibitem{turnout}
We calculate the average turnout at the regional level by using the information provided by Italian Ministry of the Interior, which surveyed it at 12pm, 7pm and 11pm of September 20$^{\text{th}}$, and at the polling stations' closing at 3pm of September 21$^{\text{st}}$.




\bibitem{conley1999}
T.G.~Conley, GMM estimation with cross sectional dependence. \textit{Journal of Econometrics} 92(1), 1-45 (1999). 

\bibitem{Qin2020}
J.~Qin, {\it et~al.\/}, Estimation of incubation period distribution of COVID-19 using disease onset forward time: a novel cross-sectional and forward follow-up study. \textit{Science advances} 6(33), eabc1202 (2020).

\bibitem{Ferrettieabb6936}
L.~Ferretti, {\it et~al.\/}, Quantifying SARS-CoV-2 transmission suggests epidemic control with digital contact tracing. \textit{Science} 368(6491), (2020).


\bibitem{Fetzer2020}
T.~Fetzer, T.~Graeber, Does Contact Tracing Work? Quasi-Experimental Evidence from an Excel Error in England. \textit{medRxiv}, (2020).

\bibitem{Monodeabe8372}
M.~Monod, {\it et~al.\/}, Age groups that sustain resurging COVID-19 epidemics in the United States. \textit{Science} eabe8372, (2021).


\end{thebibliography}

\begingroup
\renewcommand{\section}[2]{}%

\endgroup


\newpage
	
	\setcounter{section}{0}
	\renewcommand\thesection{S.\arabic{section}}

	\setcounter{subsection}{0}
	\renewcommand\thesubsection{S.\arabic{subsection}}

	\setcounter{table}{0}
	\renewcommand\thetable{S.\arabic{table}}
	
	\setcounter{equation}{0}
	\renewcommand\theequation{S.\arabic{equation}}
	
	\setcounter{figure}{0}
	\renewcommand\thefigure{S.\arabic{figure}}
	
	\setcounter{subfigure}{0}
	\renewcommand\thefigure{S.\arabic{figure}}
	
	\setcounter{footnote}{0}
	\renewcommand\thefootnote{S.\arabic{footnote}}

\section*{Supplementary material}
	\section{Supporting qualitative evidence}
\label{app:qualitative_evidence}

\begin{figure}[htbp]
		\caption{An Overview of COVID-19 Dynamics in Italy}
\label{fig:dynamics}
\centering
{\includegraphics[width=1\textwidth]{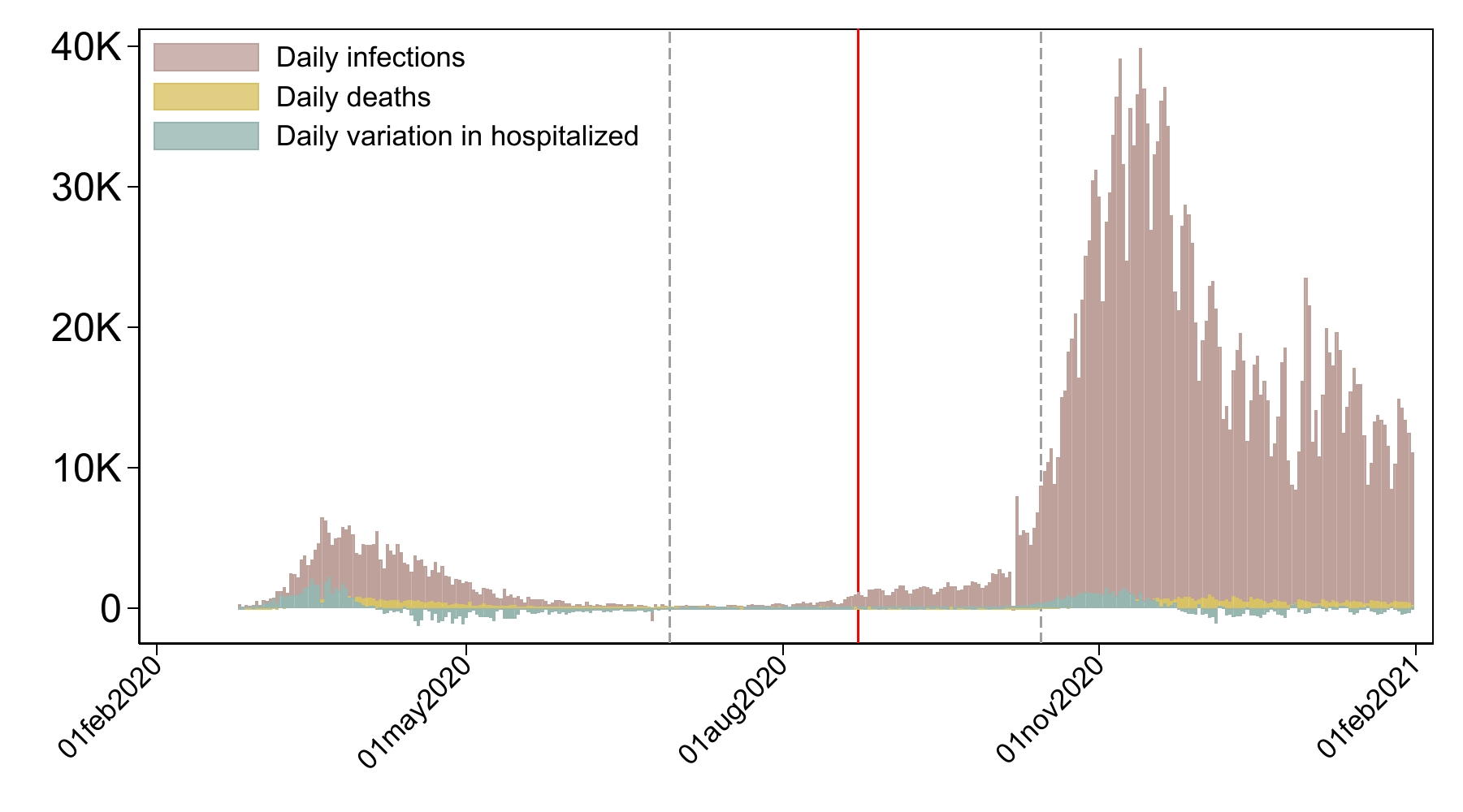}}
 \caption*{\scriptsize{\textbf{Notes}: The figure reports information on the aggregate number of individuals who tested positive to COVID-19 in Italy between February 2020 and January 2021. The dashed vertical lines represent the first and the last day included in our estimation sample. The solid vertical line represents the beginning of the campaign period used in the main specifications.}}
\end{figure}

\clearpage
\clearpage

\section{Data Appendix}
\label{app:list}

\subsection*{List of data sources (Article)}
\subsubsection*{Daily epidemiological bulletin on the spread of COVID-19 in Italy.}
(available at \url{https://github.com/pcm-dpc/COVID-19/tree/master/dati-regioni}).

We access this publicly available source to retrieve information on the daily number of infections for each region and the daily stock of individuals seeking care in ordinary hospitals and ICUs. We also retrieve the daily number of deceased individuals and the daily number of tests conducted in each region by subtracting the stock at date $t-1$ from the stock at date $t$. Lastly, we retrieve the percentage of positive tests dividing the daily number of positive tests by the daily number of tests conducted and multiplying it by 100. We restrict our attention to bulletins published between June 29$^{\textit{th}}$, 2020 and October 15$^{\textit{th}}$, 2020. For the \textit{Trentino-Alto Adige Region}, we manually compute the region-level aggregate figures because raw data contain separate information for the province of Trento and the province of Bolzano/Bozen.

We then apply the following transormations to all variables to achieve a better comparability across regions. First, 

\begin{equation}
    \widehat{y}_{r,t}=\log \Big(\frac{\tilde{y}_{r,t}\times 100,000}{Population_r}\Big),
\end{equation}
where $\tilde{y}_{r,t}$ is each of the six epidemiological outcomes of interest as originally included in the data source or as described in the previous paragraph, $Population_r$ is the region's population according to ISTAT data (see below) as of January 1$^{\textit{st}}$, 2020, and $\widehat{y}_{r,t}$ is the transformed indicator. Second, we compute a seven-day moving average as follow to get to the six outcomes $y_{r,t}$ estimated using equation 1 and equation 2. Formally, 

\begin{equation}
    y_{r,t}=\frac{\sum_{s\in[-6,0]}\widehat{y}_{r,t-s}}{7}.
\end{equation}
In turn, our final sample observes 103 daily seven-day moving averages for each of the twenty Italian regions, for a total of 2,060 observations. \\
\subsubsection*{Voter turnout on September 20$^{\textit{th}}$ and 21$^{\textit{st}}$, 2020 referendum.}
(available at \url{https://dait.interno.gov.it/elezioni/open-data/dati-referendum-20-21-settembre-2020}).

This data contains information on the number of individuals eligible to vote for the constitutional referendum in each municipality, as well as the number of individuals who decided to vote by 12 pm, 7 pm, 11 pm on September 20$^{\textit{th}}$, and by the closing time at 3 pm on September 21$^{\textit{st}}$. 

We aggregate data from all municipalities belonging to the same region to get such measures at the regional level. We then define turnout as follow,

\begin{equation}
Turnout_{r,h}=\frac{Voters_{r,h}}{Eligible_r}.
\end{equation}

\subsubsection*{Mentions of political leaders on regional TV news.} (data are available upon request).

We accessed this data by visiting the Italian public TV provider (RAI) and collect information on all the times each of the prominent political leaders in Italy have been mentioned on public TV. We search for Luigi di Maio, leader \textit{de facto} of the 5-Star Movement; Giorgia Meloni, president of \textit{Fratelli d'Italia}; Matteo Renzi, leader \textit{de facto} of \textit{Italia Viva}; Matteo Salvini, secretary of \textit{Lega}; Nicola Zingaretti, secretary of \textit{Partito Democratico}.

We then restricted our attention to mentions during the editions of local TV news (\textit{Tg Regionale}), which go on air on the third-main channel of the broadcaster daily at 2 pm; 7.30 pm; 0.10 am (on weekends at 2 pm; 7.30 pm; 11 pm).

Lastly, we summed the number of mentions of all leaders for each date and region.

\subsubsection*{Public events with Senator Matteo Salvini} (available at \url{https://legaonline.it/t_tour19.asp?l2=1992&archivio=1}).

We manually counted the number of events reported in the list for each day and region.

\subsubsection*{Google COVID-19 Community Mobility Reports} (available at \url{https://www.google.com/covid19/mobility/}).

We use daily information from each of the twenty regions relative to the variables labeled as \textit{Retail and recreation percent change from baseline}; \textit{Grocery and Pharmacy percent change from baseline}; \textit{Workplaces percent change from baseline}; \textit{Residential change from baseline}. We exclude the remaining two measures (i. e., \textit{Parks percent changes from baseline} and \textit{Transit Stations percent changes from baseline}) because of several missing observations in the period under consideration. 

\textit{Retail and recreation percent change from baseline}, \textit{Grocery and Pharmacy percent change from baseline}, and \textit{Workplaces percent change from baseline} measure the percentage change in the number of Google customers who are located by their smartphone in each of those areas. \textit{Residential change from baseline}, instead, measures the percentage change in the average number of hours that Google customers are located by their smartphone in their home.

We restrict our focus to data covering mobility between June 29$^{\textit{th}}$ and October 15$^{\textit{th}}$, 2020.

\subsection*{List of data sources (Supplementary Material)}
\subsubsection*{Municipality elections held on September 20$^{\textit{th}}$ and 21$^{\textit{st}}$, 2020}

We retrieve the list of all municipality elections that took place on the same day as the constitutional referendum and the seven regions' regional elections to calculate the share of each region's voters subject to these elections. We use data on the population eligible to vote at the municipality level described in the previous sub-section to calculate the shares. 

\subsubsection*{Geographic, socio-economic, and demographic characteristics from ISTAT}

In Table \ref{tab:table_results_region_controls}, we propose a specification in which regional-by-week linear pre-treatment trends are replaced by weekly pre-treatment trends in observable characteristics that might explain the spread of COVID-19. These characteristics include (a) the share of voters subject to municipality elections just described;  (b) the region's surface; (c) population density; (d) percentage of hospitalized individuals coming from another region; (e) unemployment rate; (f) employment rate; (g) employment rate among senior ($>$54 years old) individuals; (h) percentage of elderly in care; (i) percentage of population covered by ultra-broadband internet connection; (j) average travel times from central urban areas; (k) the average number of daily tourists per 100 residents; (l) percent of municipalities providing childcare services; (m) mountainous surface; coastal surface; (n) number of museums visitors per km$^{\textit{2}}$; (o) per capita export volume; (p) an indicator equal to 1 if the region is located in Southern Italy. We collected all the latter variables from the Italian National Institute of Statistics (ISTAT). Table \ref{tab:table_descriptives_boundary} reports the descriptive statistics of these and all other variables employed in this analysis.

\begin{table}[h!]
\centering
\resizebox{.7\linewidth}{!}{
\begin{threeparttable}
\footnotesize
\caption{Descriptive Statistics}\label{tab:table_descriptives_boundary}
\begin{tabular}{lccc}
\hline\hline
&(1)&(2)&(3) \\
&Mean&St. Dev.&Obs. \\
\hline 
\multicolumn{4}{c}{(a) Measures of diffusion} \\
Cases&0.60&0.45&2,060 \\
Tests&4.44&0.52&2,060 \\
\% Positive&0.594&0.350&2,060 \\
Hospitalized&0.85&0.49&2,060 \\
ICU&0.12&0.12&2,060 \\
Deaths &0.01&0.02&2,060 \\
\hline
\multicolumn{4}{c}{(b) Regional characteristics} \\
Population ($\times$ 100K)&30.12&25.40&20 \\
Surface (km$^{\textit{2}}$)&17,398&9,712&20 \\
Density (inh./km$^{\textit{2}}$)&317&368&20 \\
\% Healthcare regional migration&8.92&6.20&20 \\
Unemployment rate (\%)&10.47&5.19&20 \\
Employment rate (\%)&59.38&10.20&20 \\
Senior Employment rate (\%)&55.15&6.64&20 \\
Elderly care patients (\%)&1.24&0.92&20 \\
Fiber Internet connection (\%)&14.50&2.97&20 \\
Travel times from urban areas (mins.)&51.45&10.60&20 \\
Daily tourists (per 100 inh.)&9.59&10.83&20 \\
Childcare provision (\%)&58.20&23.91&20 \\
Mountainous surface (km$^{\textit{2}}$)&2.59&1.23&20 \\
Coastal surface (km$^{\textit{2}}$)&0.18&0.22&20 \\
No. Museums (per inh.)&0.01&0.01&20 \\
Export volume (per inh.)&44,066&45,580&20 \\
South&0.45&0.51&20 \\
\hline
\multicolumn{4}{c}{(c) Election characteristics} \\
Campaign&0.35&0.49&20 \\
Share municipality election&0.08&0.11&20 \\
Voter turnout (Sunday at 12PM)&0.12&0.03&20 \\
Voter turnout (Sunday at 19PM)&0.30&0.08&20 \\
Voter turnout (Sunday at 11PM)&0.40&0.09&20 \\
Voter turnout (Overall)&0.55&0.11&20 \\
\hline\hline
\end{tabular}

\caption*{\scriptsize{\textbf{Notes}: The table show the mean, the standard deviation and the number of observations for each vaiable employed in the analysis. The summary statistics in panel (a) are calculated considering the entire estimation sample, while those showed in panels (b) and (c) refer to the regional value of the characteristics as in 2019 and during the election days, respectively.}}
\end{threeparttable}
}
\end{table}

\clearpage

\section{Robustness tests}
\label{app:robustness}

Besides showing that the six epidemiological outcomes of interest were following the same dynamics in treated and control regions before the start of the campaign (see Figure 3), we performed various robustness tests to validate our estimates and research design further. 

First, in Table S.2, we replace the linear pre-treatment weekly regional trends with weekly trends in several observable characteristics that can affect the spread of COVID-19. The variables included in the analysis are defined and reported in Table S.1. On the one hand, this exercise allows establishing that this paper's main results do not rely on the control for pre-trends in the outcomes. On the other hand, the exercise is beneficial to exclude that observed patterns are due to differences in other dimensions, which can confound our results. In particular, this robustness check excludes the possibility that our results depend -- among others -- on differences in touristic activities, availability of smart-working solutions, the share of elderly individuals who are in care, diffusion of pre-school services, and commuting times. 

Second, in Figure S.2, we show that our main results do not depend on the specific choice of considering all weeks after week 34 as post-treatment and the weeks before as pre-treatment. In each iteration, we estimate a version of equation 1 in which the $Post_t$ indicator is equal to 1 starting from the week specified on the horizontal axis and 0 otherwise. Week 30 was the week when the national government announced the election dates, while week 38 is the week during which the election took place. This exercise allows us to establish that our results do not depend on our definition of the campaign period. More specifically, it might be the case that party leaders and candidates start a campaign before the official start reported in the law, which is 30 days before the election. Our results show that the choice to consider week 34 -- as in the law -- in the main specification is not driving our results. Indeed, the estimates are qualitatively equal if we allow the campaign period to begin during week 33 or 32 (i.e., up to 45 days before the election). 

Third, in Figure S.3, we show that our results do not rely on a specific region's presence in the sample. In each iteration, we estimate equation 1 by removing from the sample the region specified on the horizontal axis. We find that the estimated coefficients are very stable, excluding the possibility that our estimates depend on one outlier region, which either has low diffusion of COVID-19 and no regional election, or high spread of the infection and the presence of regional elections. 

Lastly, we assess the robustness of our statistical inference. The Conley HAC standard errors \textit{(19)} account both for spatial correlation of units that are geographically close and for auto-correlation over time. However, they come with the limitation that the choice of cutoffs is arbitrary. In Figure S.4, we show that our analysis's statistical inference bases on very conservative cutoffs. More specifically, at each iteration, we estimate equation 1 by imposing a non-zero correlation among observations within the distance cutoff specified on the horizontal axis and the lag cutoff specified on the vertical axis, and we report the p-values for the $Post_t \times Campaign_r$ indicator. For all epidemiological outcomes, we find that our main results draw statistical inference using the largest feasible p-values.  

\begin{table}[htbp]
\centering
\resizebox{1\linewidth}{!}{
\begin{threeparttable}
\normalsize
\caption{Robustness Test: Controlling for Trends in Observables}\label{tab:table_results_region_controls}
\begin{tabular}{lcccccc} \hline\hline
  & (1) & (2) & (3) & (4) &(5) & (6)   \\
& Infections & Tests &  \% Posit  & Hospit.  &  ICUs & Deaths  \\ 
 \hline
 &  &  &  & &  &   \\
 Campaign $\times$ Post & 0.070*&-0.083***&0.146***&0.240***&0.053**&0.006* \\
 & (0.033)&(0.019)&(0.031)&(0.052)&(0.020)&(0.002)\\
 &  &  &  &   &  &    \\
Observations &2,060&2,060&2,060&2,060&2,060&2,060\\
Baseline (value)&0.315&83.578&0.404&0.904&0.032&0.012\\
Region  FE &\checkmark&\checkmark&\checkmark&\checkmark &\checkmark&\checkmark  \\
Reg. charact. pre-trends & \checkmark&\checkmark&\checkmark&\checkmark&\checkmark&\checkmark \\
 Date $\times$ Adjacent reg. FE &\checkmark&\checkmark&\checkmark&\checkmark&\checkmark&\checkmark\\
 \hline\hline
\end{tabular}

\caption*{\scriptsize{Notes: The table shows the differential impact of electoral campaign between regions voting both for the referendum and the regional election (treatment) and those only voting for the referendum (control), as obtained through the estimation of Equation 1 showed in the article. 
All dependent variables are the natural logarithm of the seven-day moving average of the daily number at the regional level. See Section 3 for details. $Campaign$ is a dummy equal to 1 for treatment regions, and 0 otherwise. $Post$ is a dummy equal to 1 for all the dates since August 21st, and 0 otherwise.  All columns include region fixed effects, pre-treatment trends in region-specific characteristics and date-adjacent regions fixed effects. Standard errors are robust to serial and spatial correlation and computed imposing maximum cutoff for both spatial and time lags. $\dagger$,  *,**,*** represent the 10\%, 5\%, 1\%, 0.1\% significance levels, respectively.}}
\end{threeparttable}
}
\end{table}

\clearpage

\begin{figure}[h!]

\centering
\caption{Robustness Test: Start of the Electoral Campaign}\label{fig:figure_campaign_band}

\subcaptionbox{: New infections} {\includegraphics[width=0.4\textwidth]{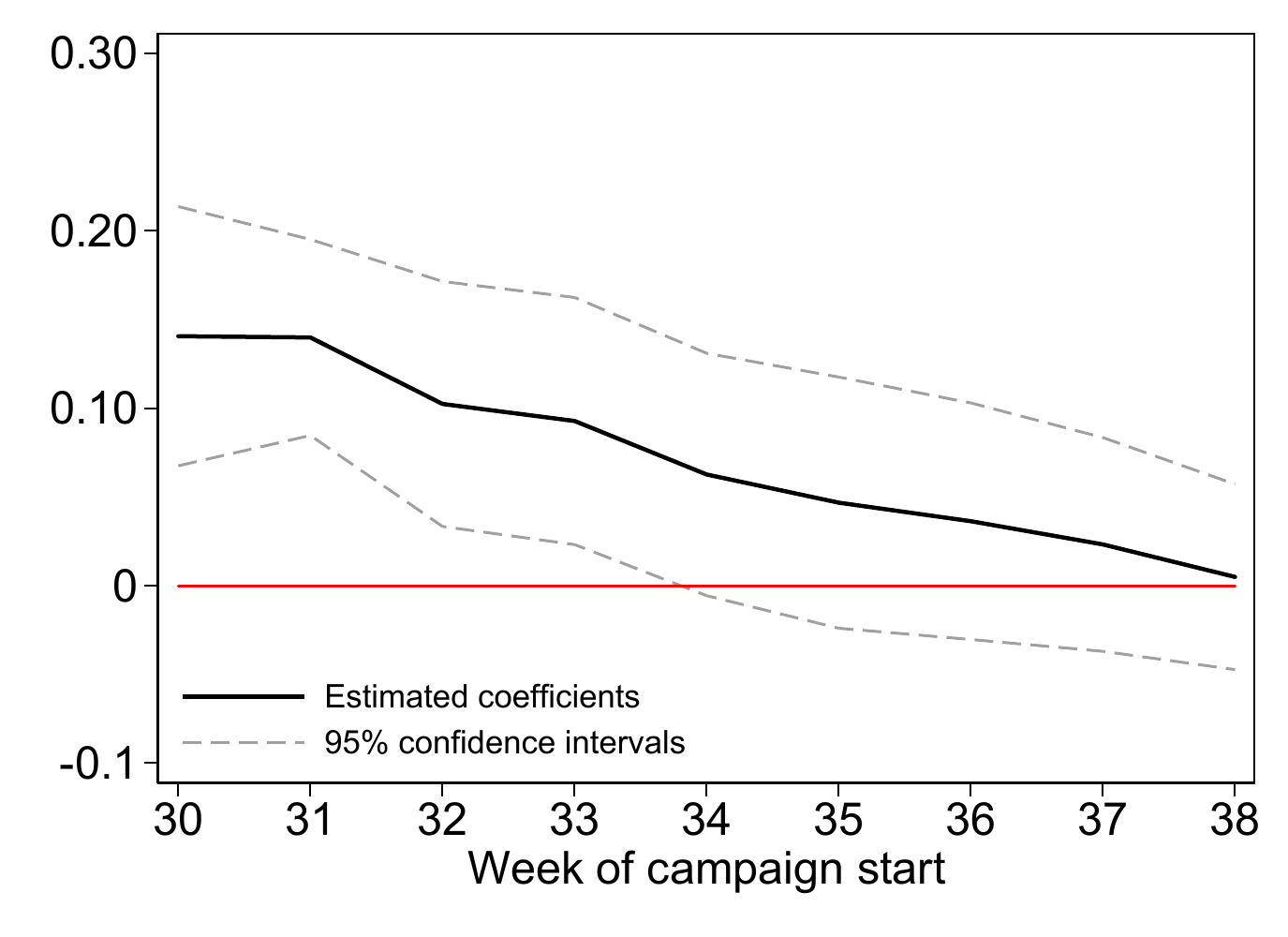}}
\hspace{1cm}
\subcaptionbox{: Tests} {\includegraphics[width=0.4\textwidth]{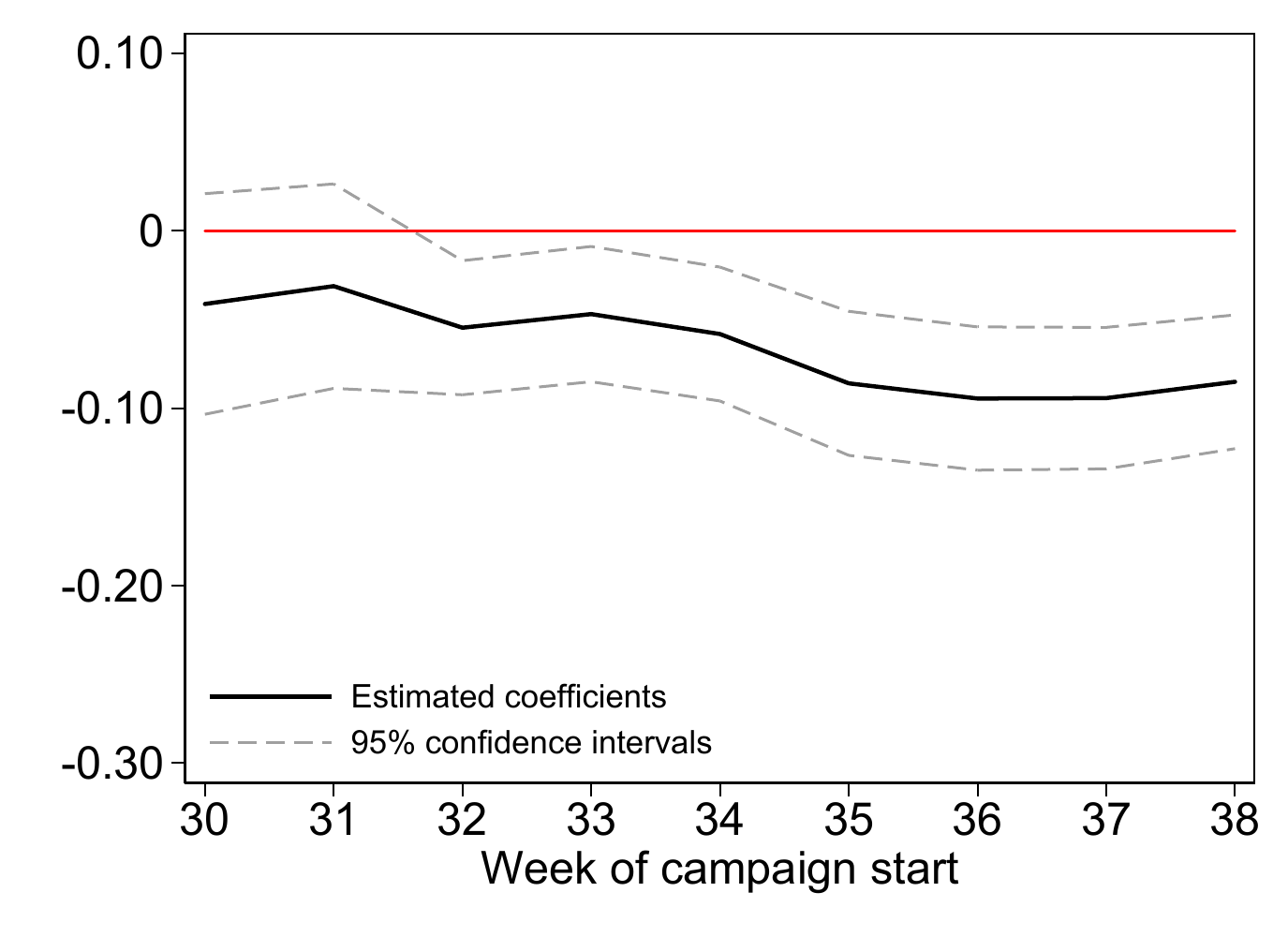}}
\vspace{.01cm}

\subcaptionbox{: \% Positives} {\includegraphics[width=0.4\textwidth]{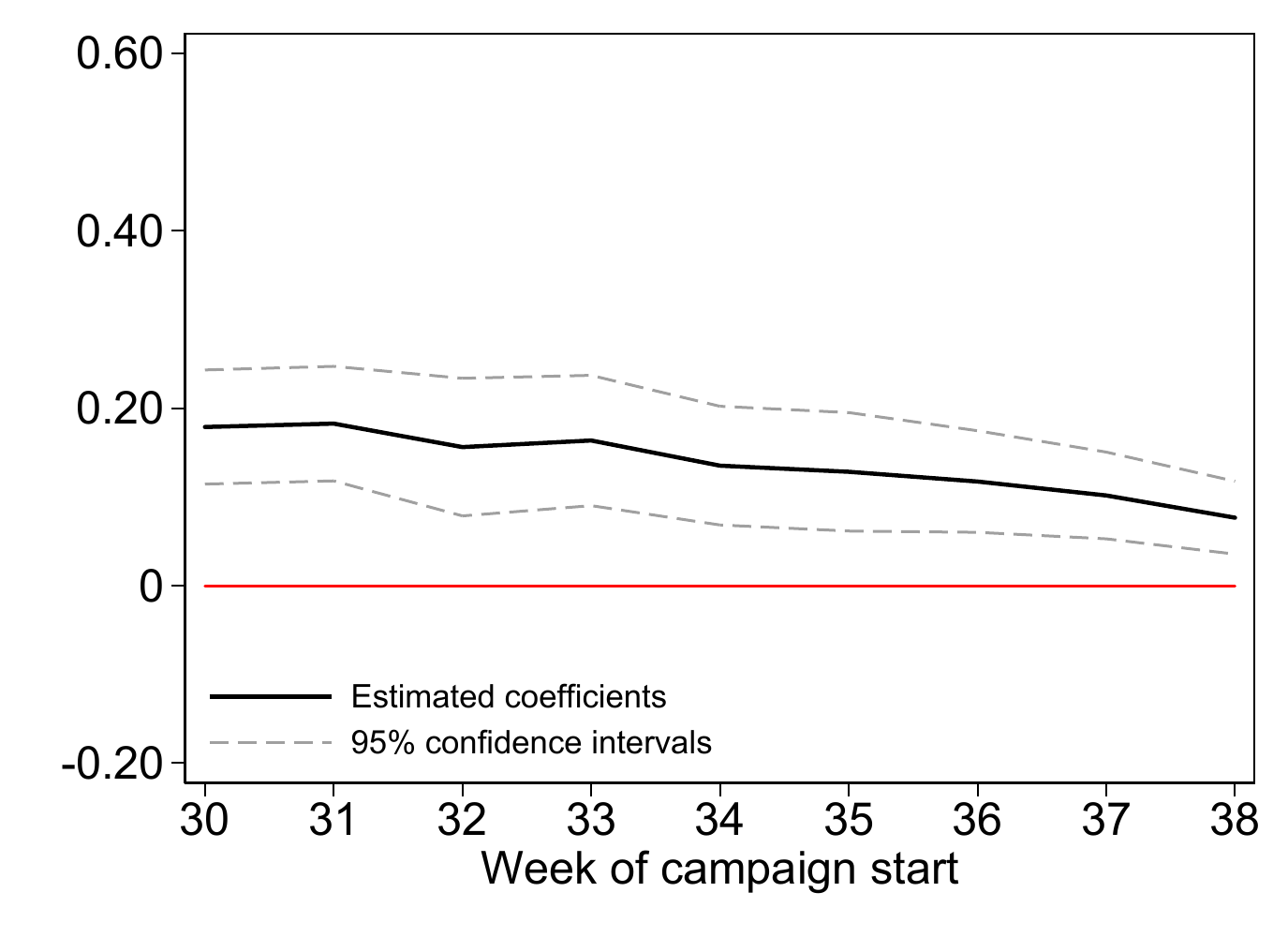}}
\hspace{1cm}
\subcaptionbox{: Hospitalized} {\includegraphics[width=0.4\textwidth]{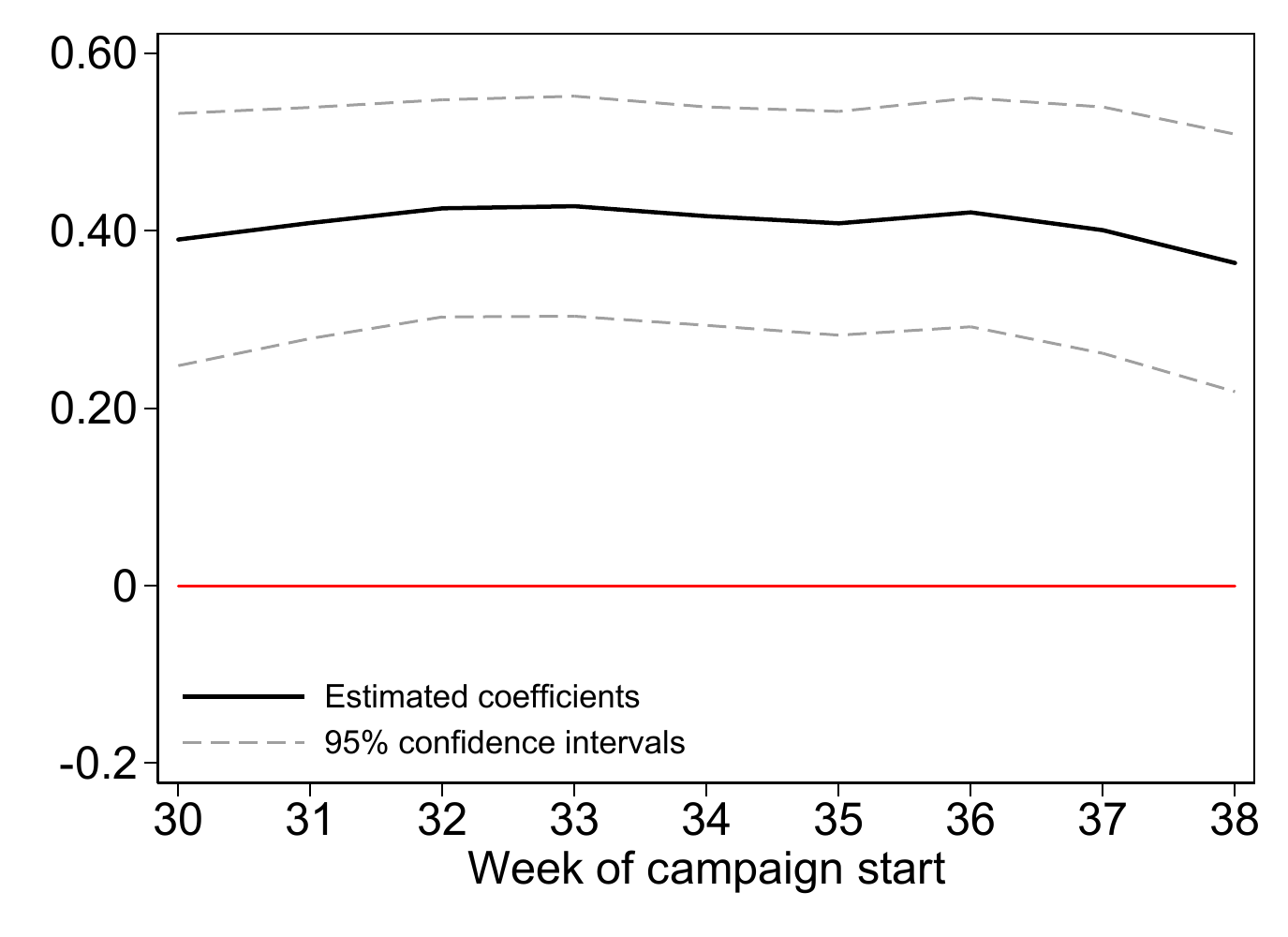}}

\vspace{.01cm}
\subcaptionbox{: ICUs} {\includegraphics[width=0.4\textwidth]{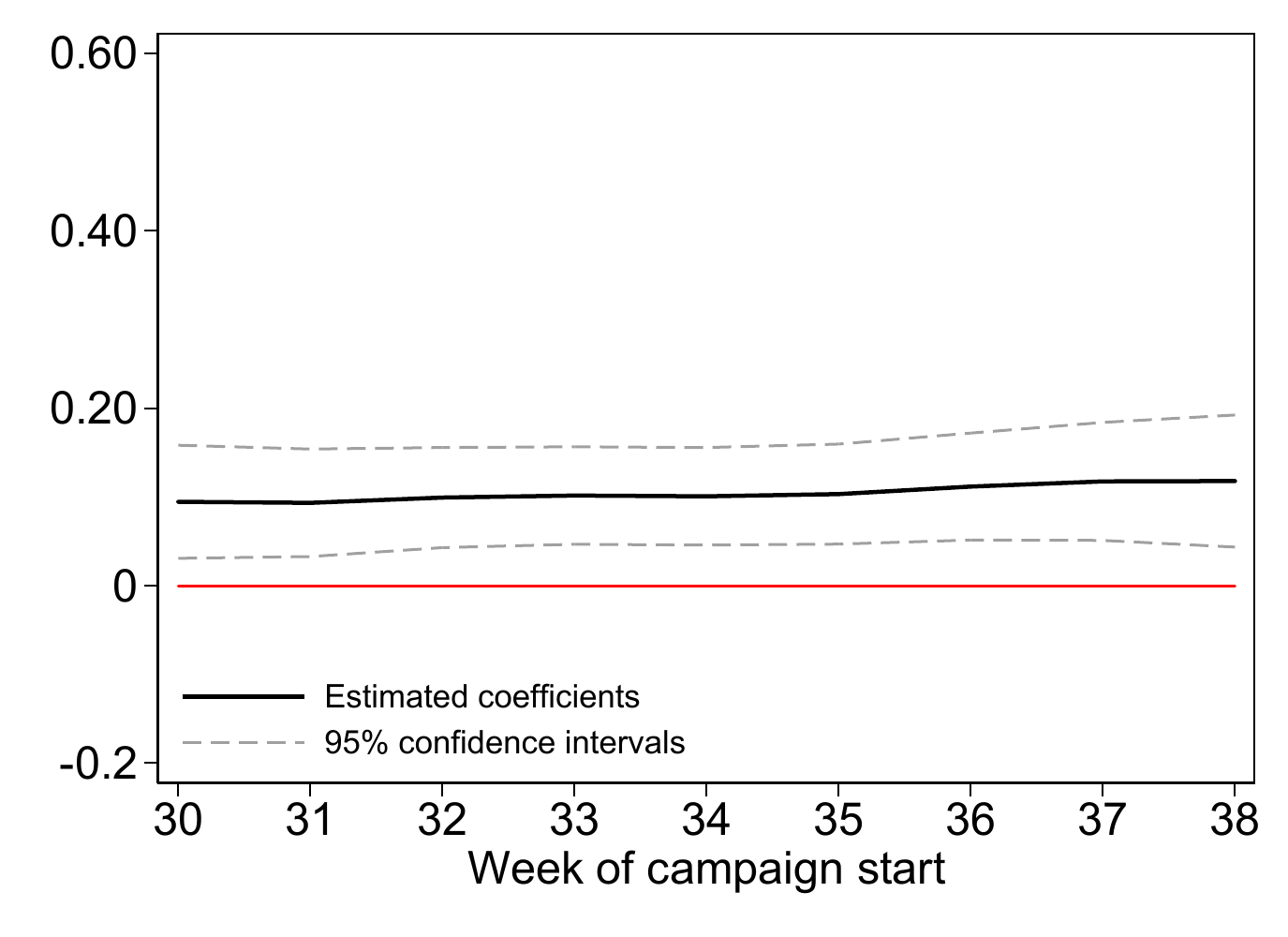}}
\hspace{1cm}
\subcaptionbox{: Deaths} {\includegraphics[width=0.4\textwidth]{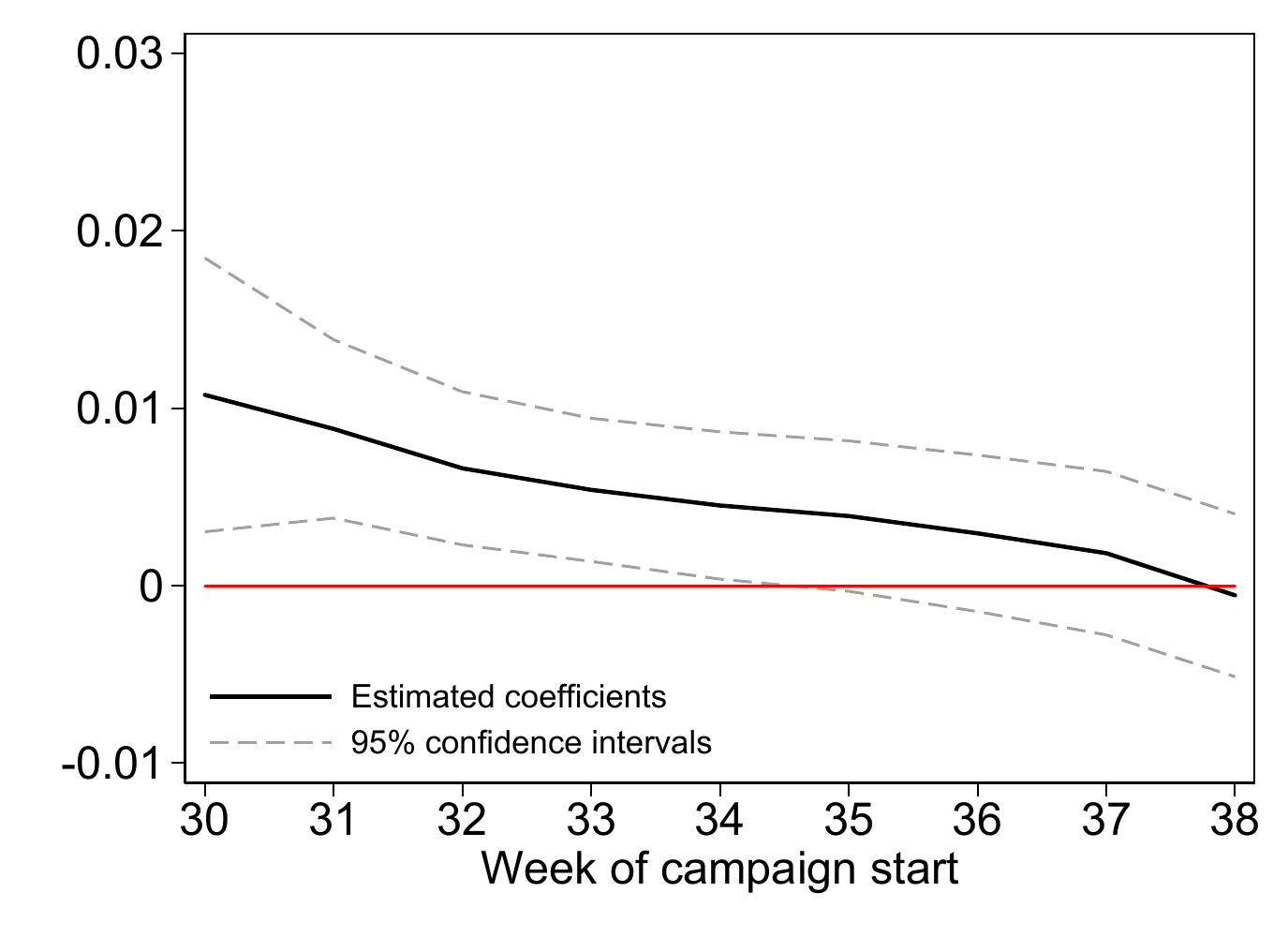}}
 
\caption*{\scriptsize{\textbf{Notes}:
The figures shows the differential impact of electoral campaign between regions voting both for the referendum and the regional election (treatment) and those only voting for the referendum (control), as obtained through the estimation of Equation 1 showed in the article, for different definitions of the starting week of the electoral campaign (horizontal axis), and together with the 95\% confidence intervals (dashed lines). All dependent variables are the natural logarithm of the seven-day moving average of the daily number at the regional level. See Section 3 of the article for details. $Campaign$ is a dummy equal to 1 for treatment regions, and 0 otherwise. Confidence intervals are based on standard errors that are robust to serial and spatial correlation and computed imposing maximum cutoff for both spatial and time lags.}} 
\end{figure}

\begin{figure}[h!]
\centering
\caption{Robustness Test: Dropping one Region at the Time}\label{fig:figure_drop_region}

\subcaptionbox{: New infections} {\includegraphics[width=0.4\textwidth]{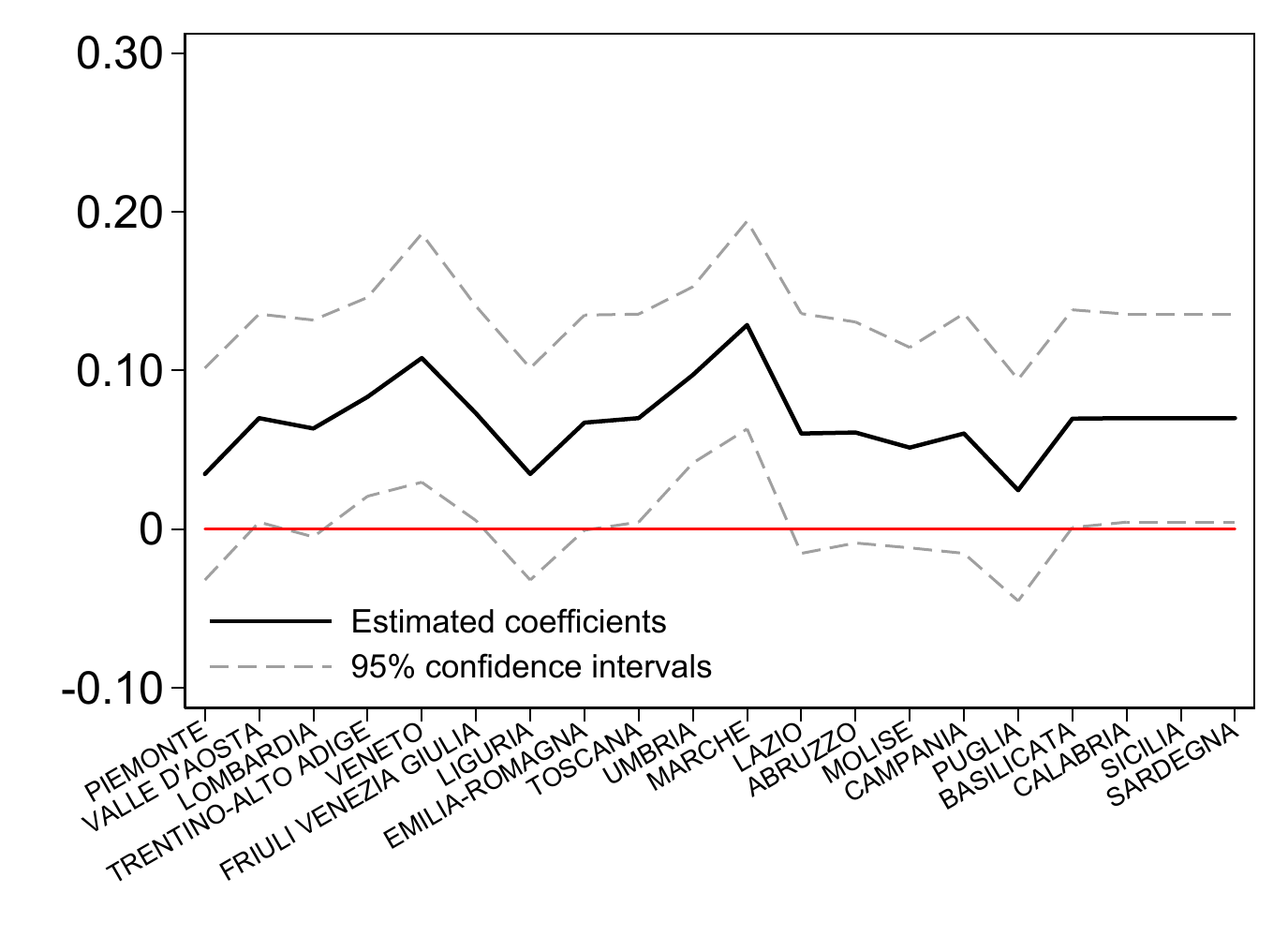}}
\hspace{1cm}
\subcaptionbox{: Tests} {\includegraphics[width=0.4\textwidth]{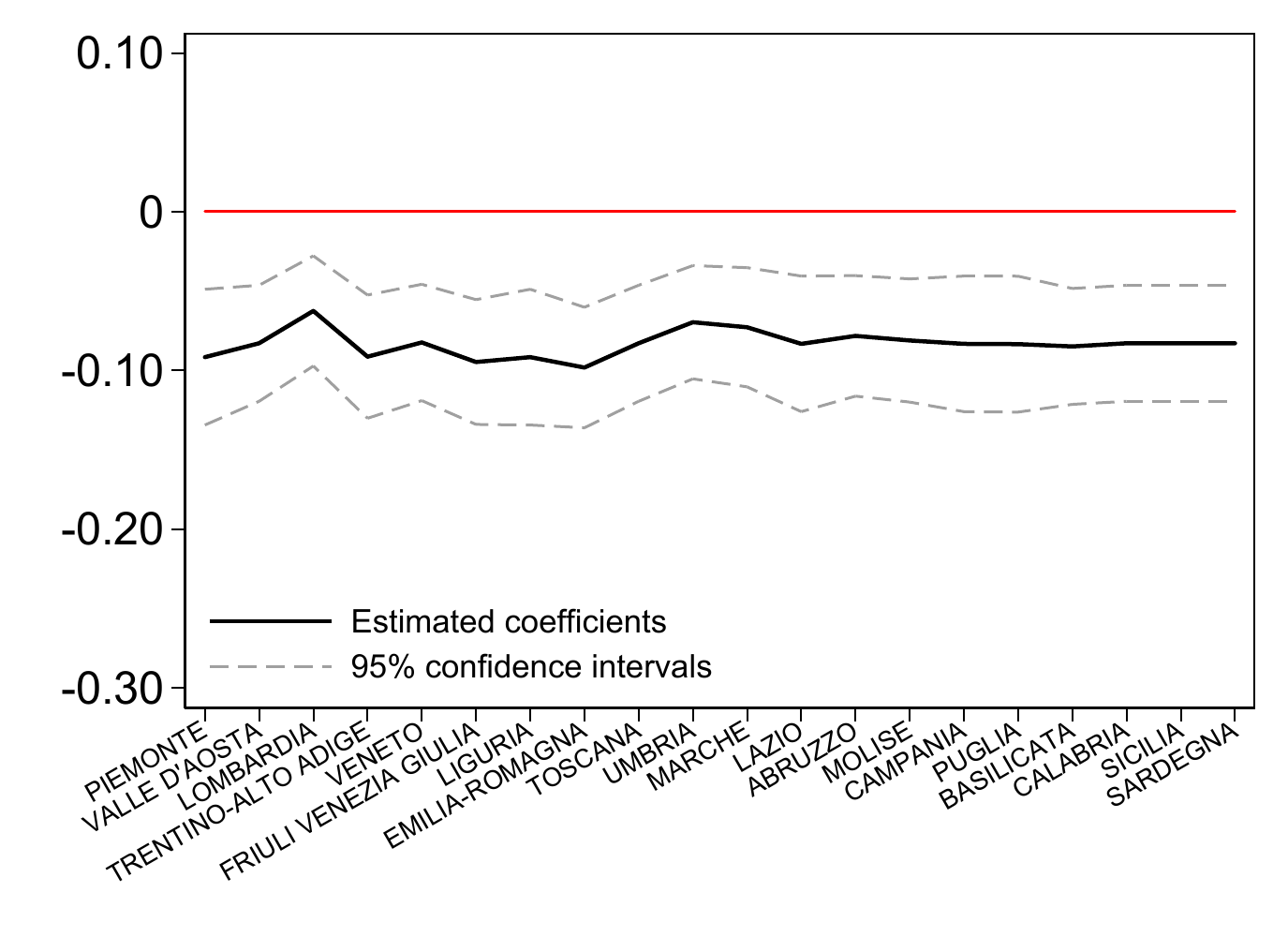}}
\vspace{.01cm}

\subcaptionbox{: \% Positives} {\includegraphics[width=0.4\textwidth]{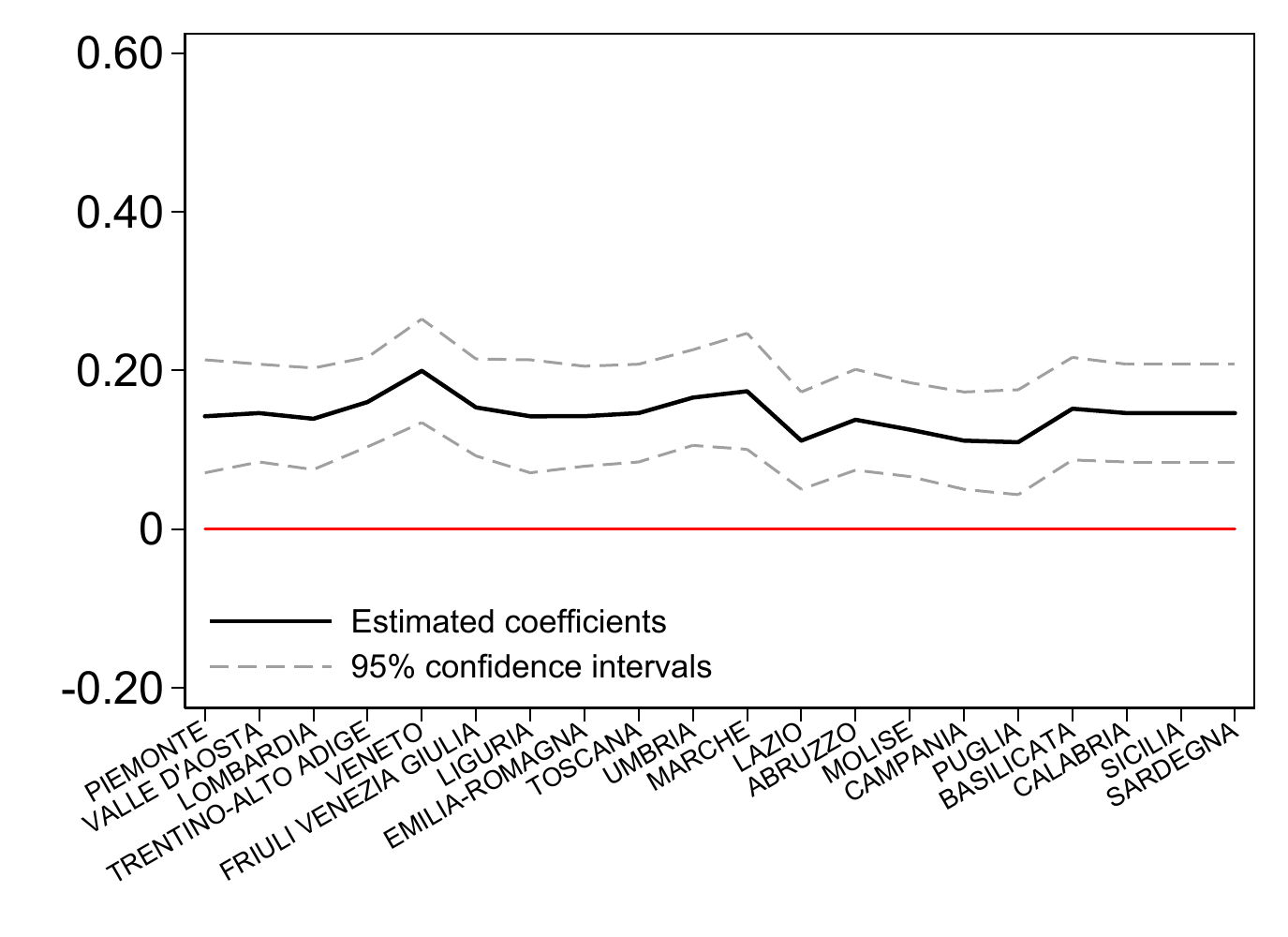}}
\hspace{1cm}
\subcaptionbox{: Hospitalized} {\includegraphics[width=0.4\textwidth]{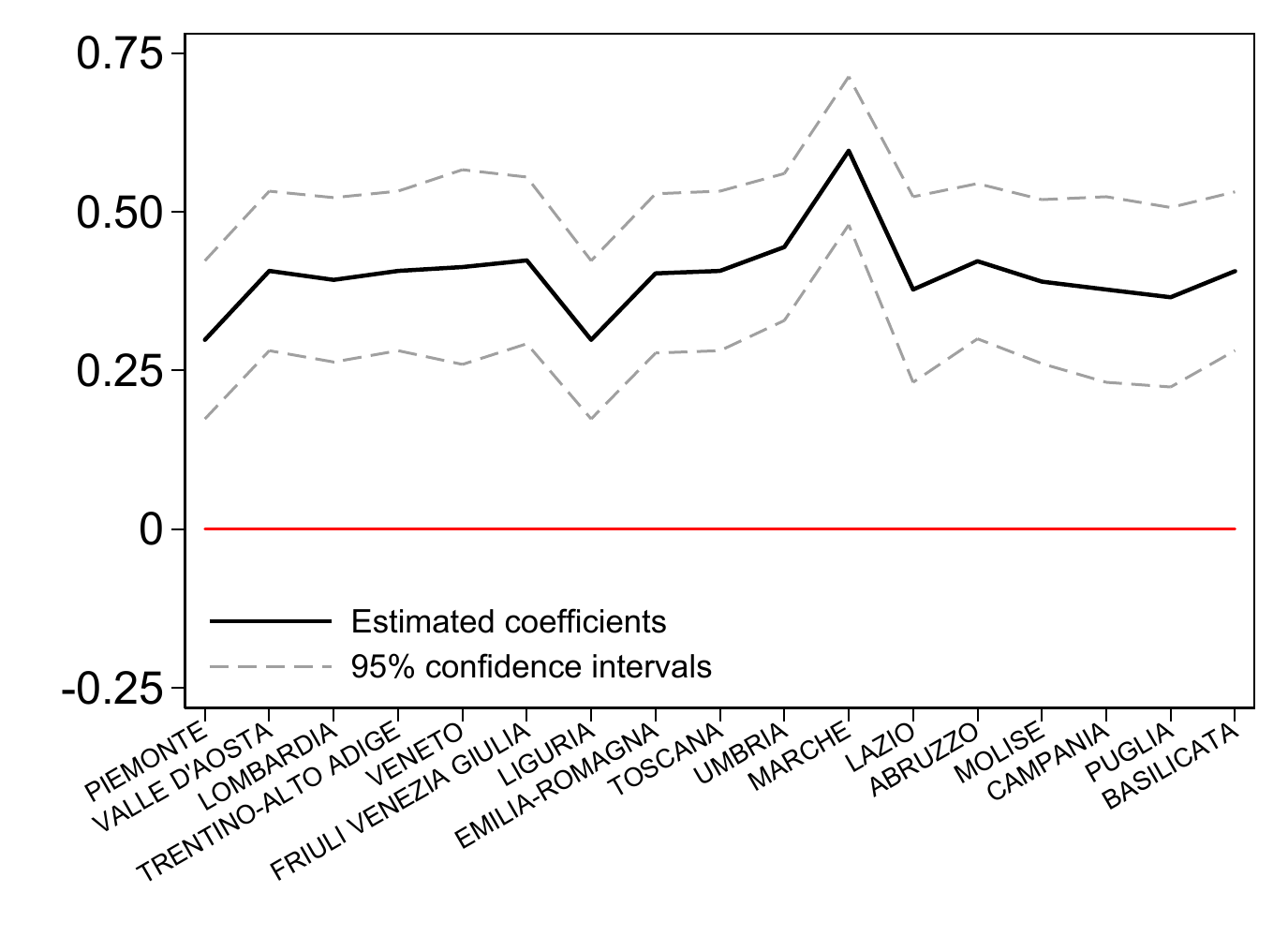}}

\vspace{.01cm}
\subcaptionbox{: ICUs} {\includegraphics[width=0.4\textwidth]{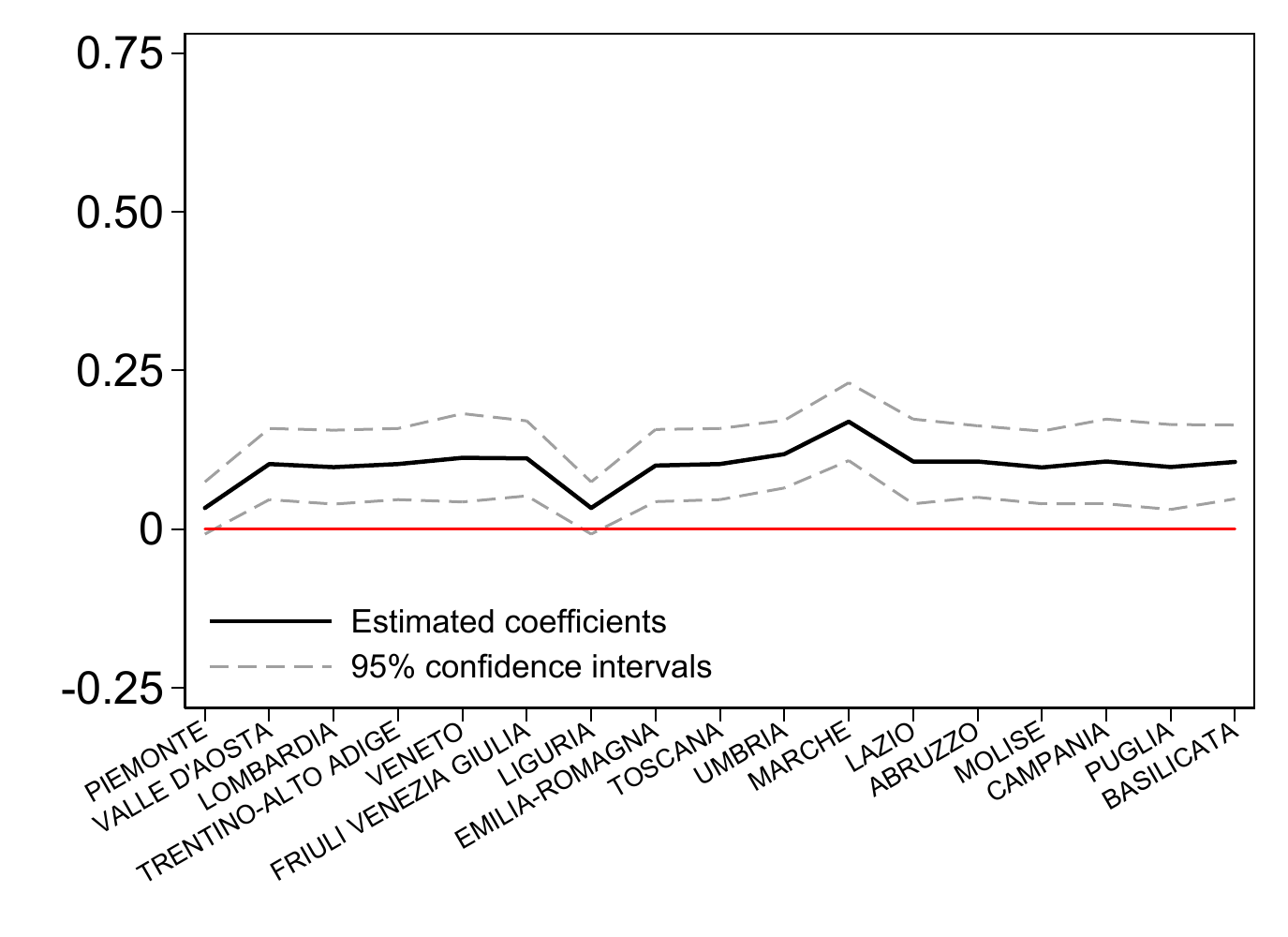}}
\hspace{1cm}
\subcaptionbox{: Deaths} {\includegraphics[width=0.4\textwidth]{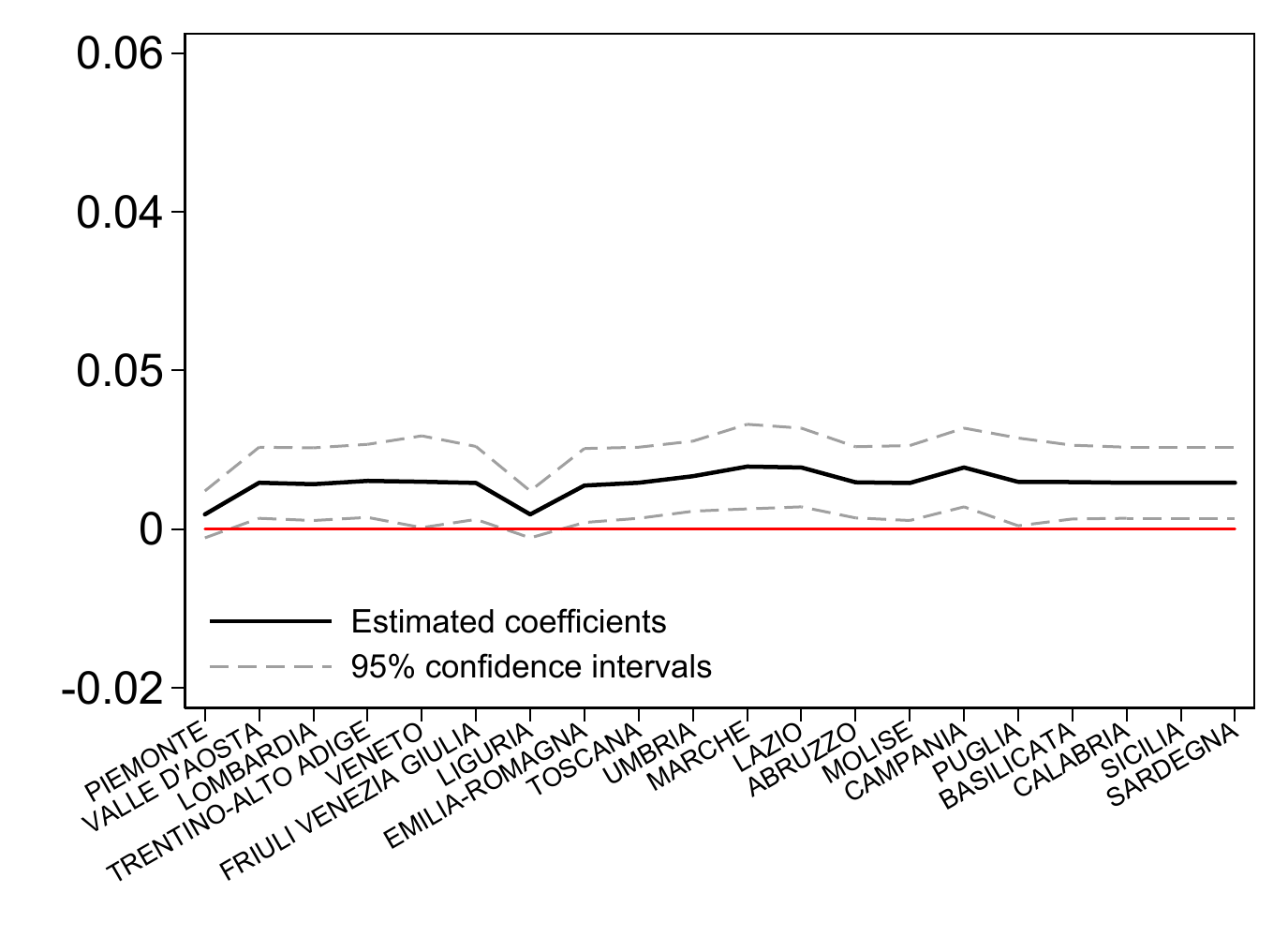}}
 
\caption*{\scriptsize{\textbf{Notes}: The figures shows the differential impact of electoral campaign between regions voting both for the referendum and the regional election (treatment) and those only voting for the referendum (control), as obtained through the estimation of Equation 1 showed in the article, for different sub-samples of observations from which the region specified on the horizontal axis has been removed, and together with the 95\% confidence intervals (dashed lines). All dependent variables are the natural logarithm of the seven-day moving average of the daily number at the regional level. See Section 3 of the article for details. $Campaign$ is a dummy equal to 1 for treatment regions, and 0 otherwise. Confidence intervals are based on standard errors that are robust to serial and spatial correlation and computed imposing maximum cutoff for both spatial and time lags.}} 
\end{figure}

\begin{figure}[h!]
\centering
\caption{Robustness Test: Conley HAC Standard Errors}\label{fig:figure_hac}

\subcaptionbox{: New infections} {\includegraphics[width=0.4\textwidth]{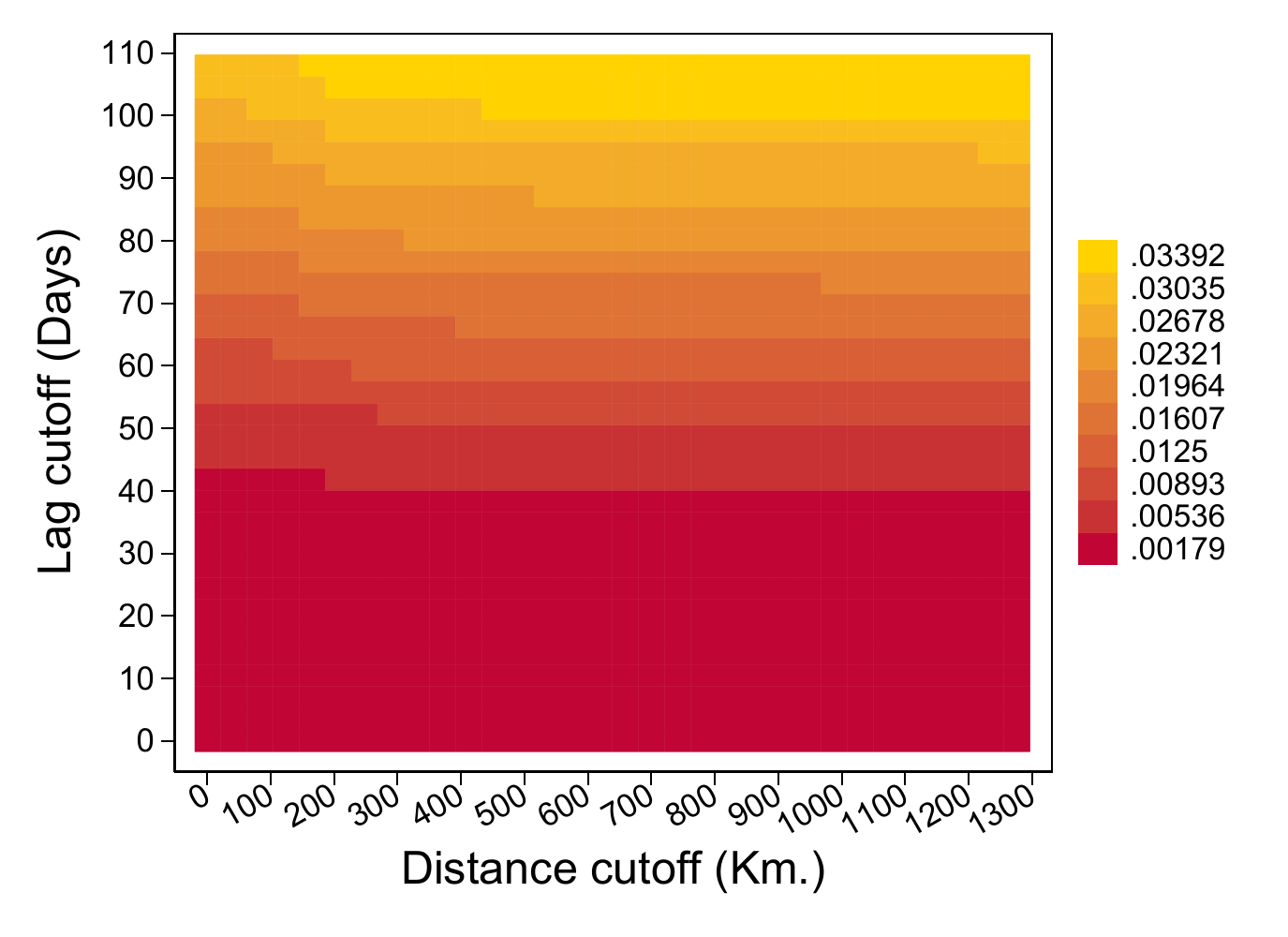}}
\hspace{1cm}
\subcaptionbox{: Tests} {\includegraphics[width=0.4\textwidth]{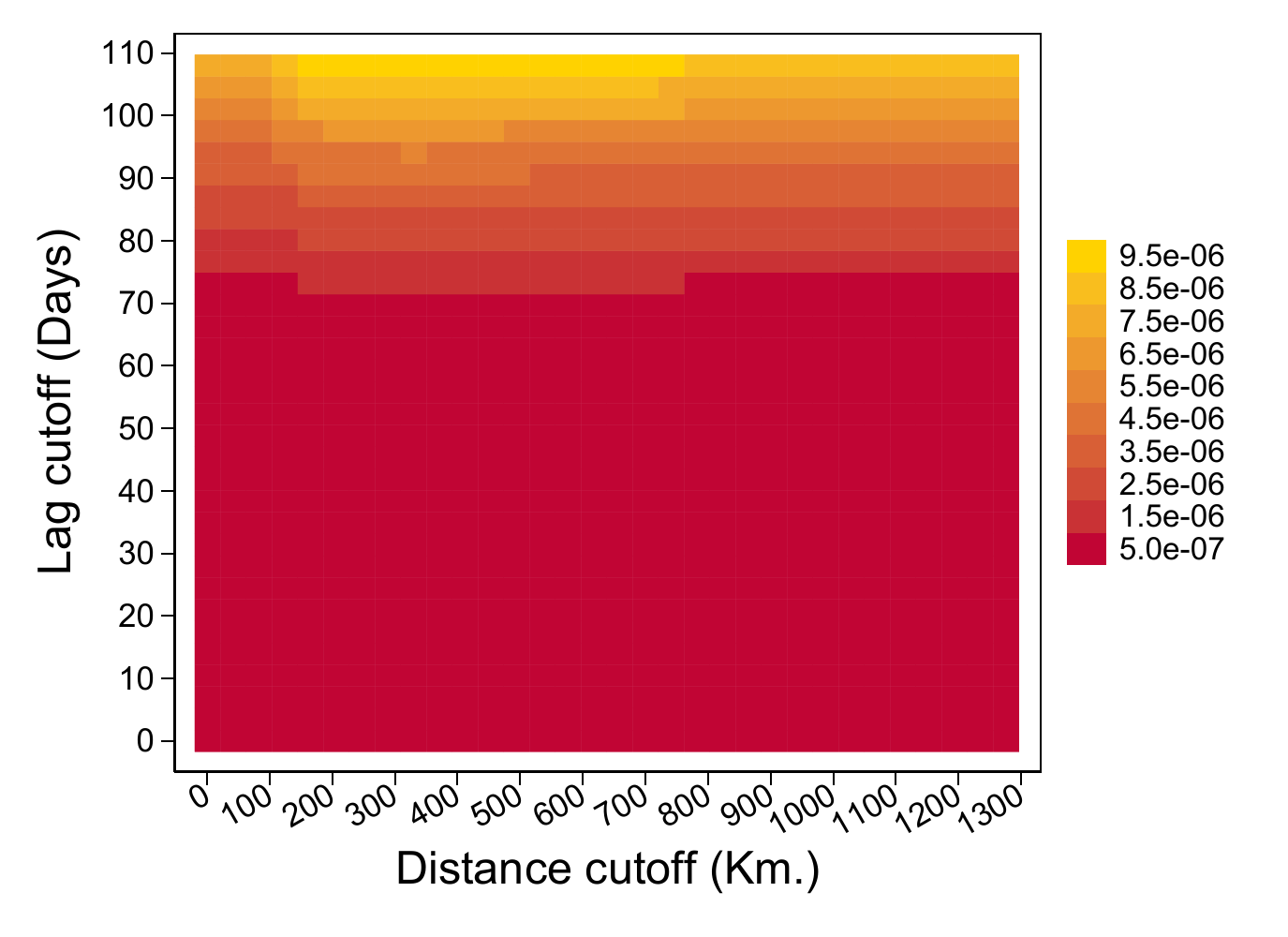}}
\vspace{.01cm}

\subcaptionbox{: \% Positives} {\includegraphics[width=0.4\textwidth]{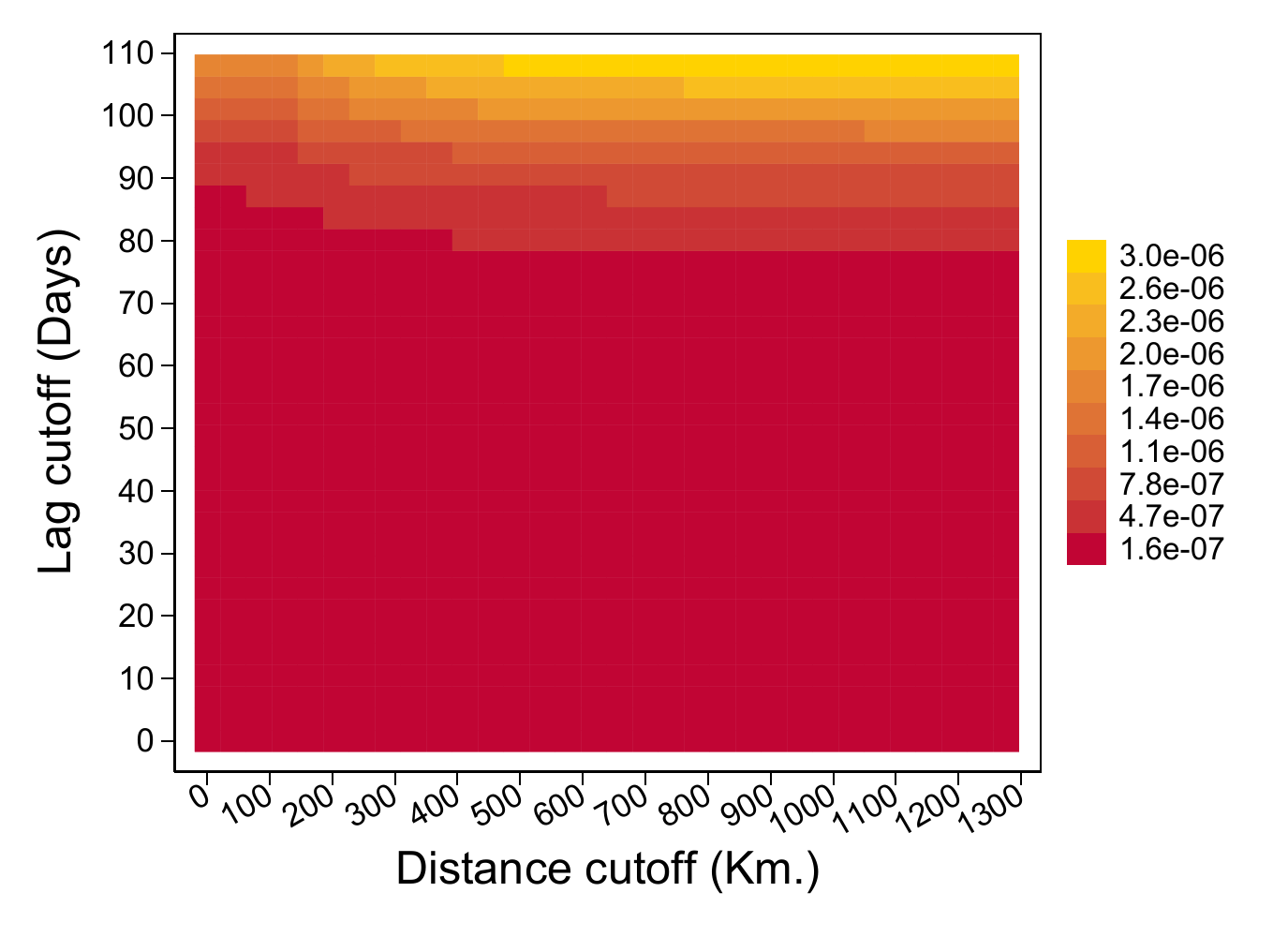}}
\hspace{1cm}
\subcaptionbox{: Hospitalized} {\includegraphics[width=0.4\textwidth]{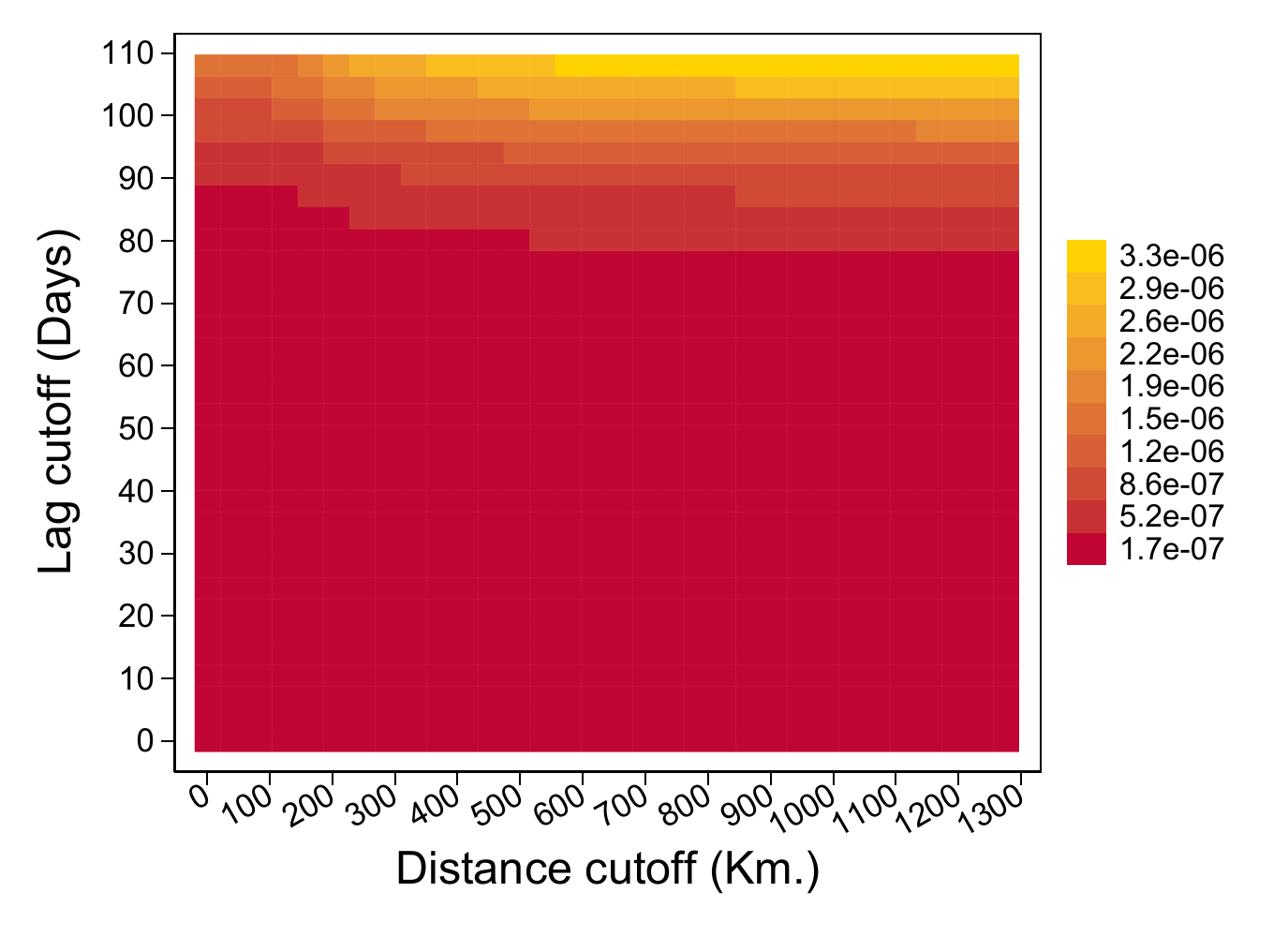}}

\vspace{.01cm}
\subcaptionbox{: ICUs} {\includegraphics[width=0.4\textwidth]{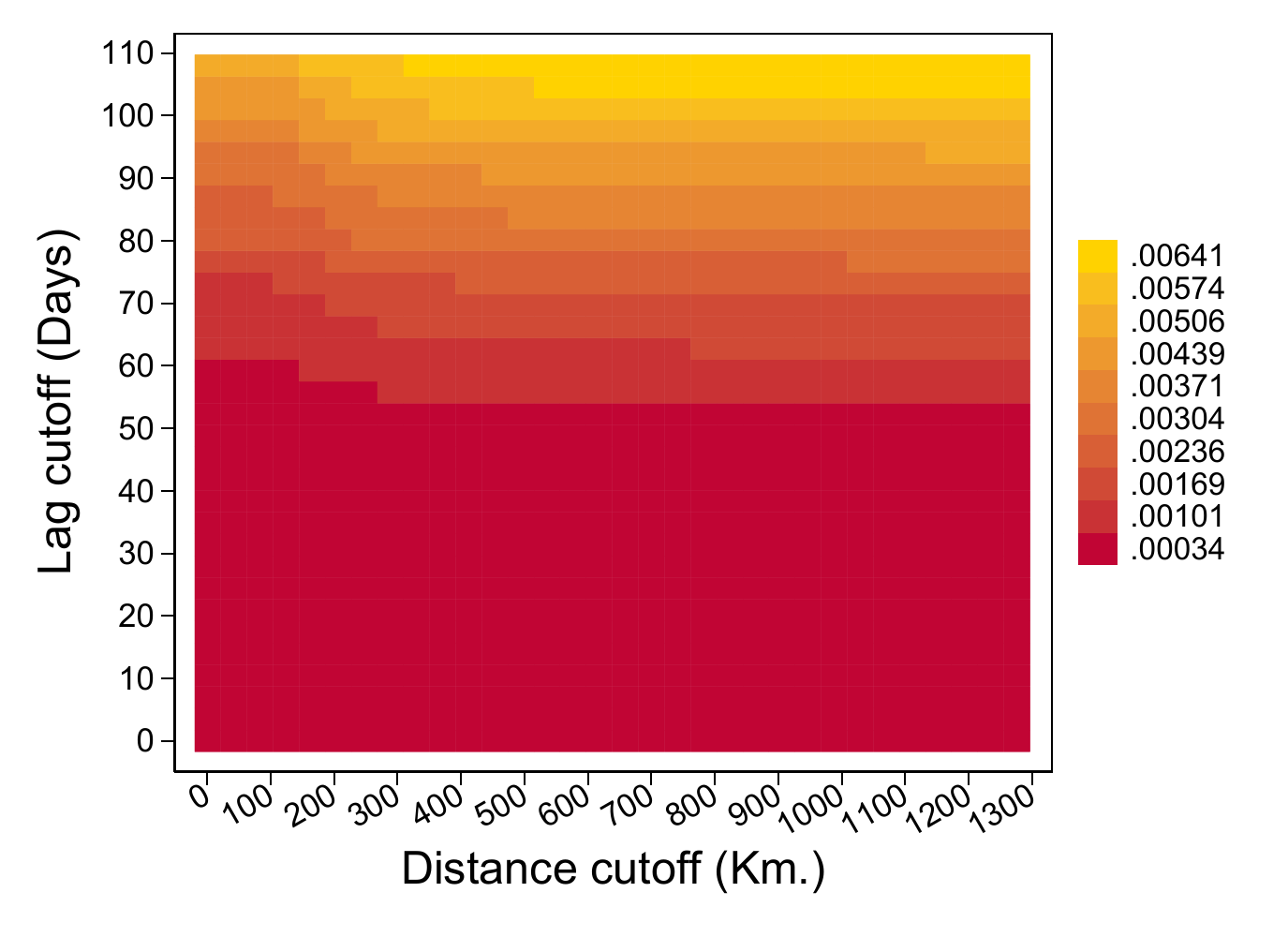}}
\hspace{1cm}
\subcaptionbox{: Deaths} {\includegraphics[width=0.4\textwidth]{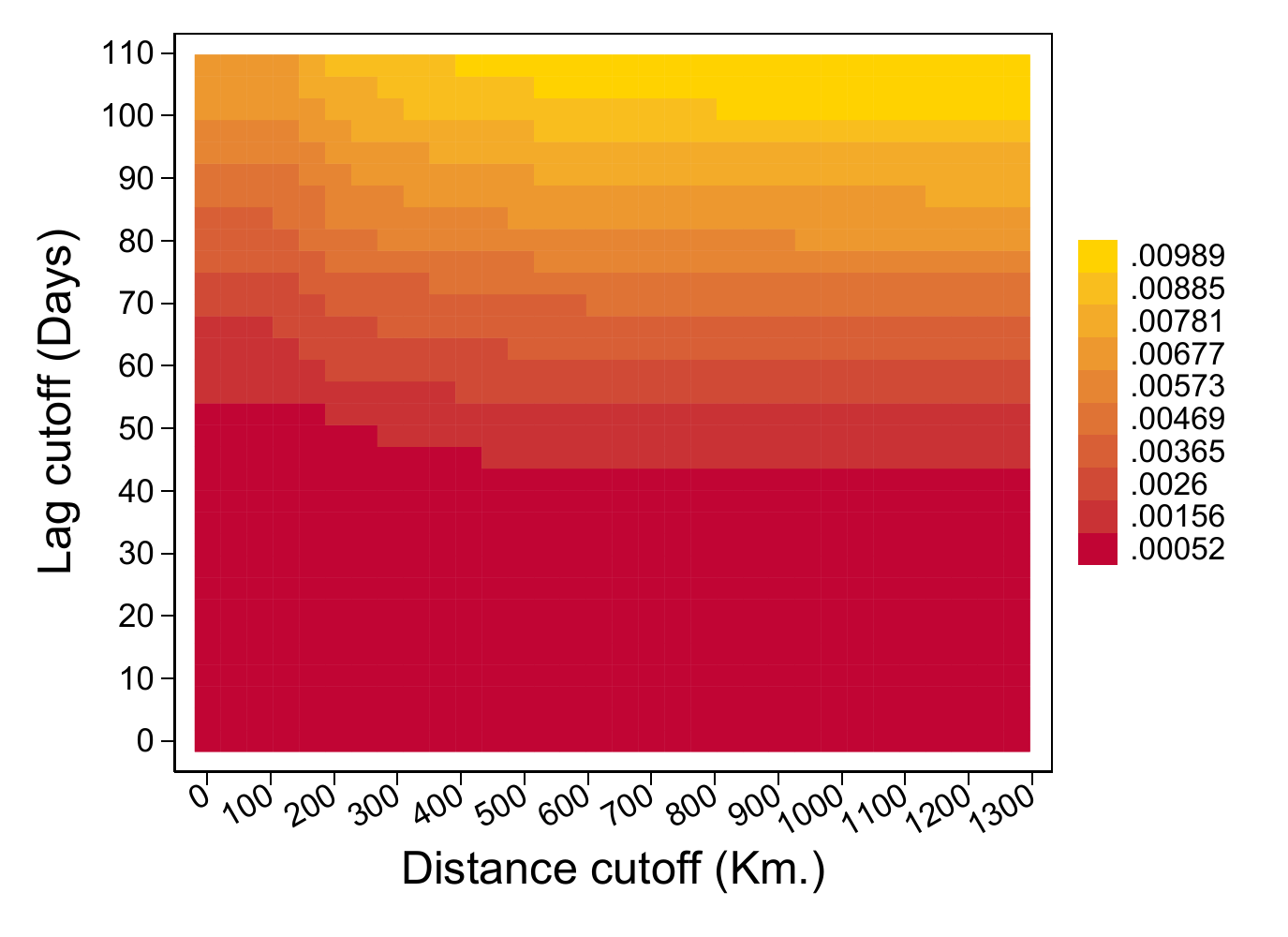}}
 
\caption*{\scriptsize{\textbf{Notes}: The figure reports p-values from estimating Equation 1 showed in the article in multiple regressions using Conley HAC standard errors \textit{(19)} with time and spatial lags vary from their minimum to their maximum feasible value. Each iteration represents and increase to the distance cutoff equal to 25 km. or an increase in the time cutoff equal to 3 days. All dependent variables are the natural logarithm of the seven-day moving average of the daily number at the regional level. See Section 3 of the article for details. $Campaign$ is a dummy equal to 1 for treatment regions, and 0 otherwise. Confidence intervals are based on standard errors that are robust to serial and spatial correlation and computed imposing maximum cutoff for both spatial and time lags.}} 
\end{figure}

\end{document}